# Terahertz Generation and Detection through Gain-Enhanced Interband Photomixing in Quantum Well Structures

Yifan Zhao[1,2], Shahed-E- Zumrat[1,2], Szu-An Tsao[1,2], Mona Jarrahi[1,2*]

[1]Electrical and Computer Engineering Department, University of California; Los Angeles, 90095, USA.

[2]California NanoSystems Institute, University of California; Los Angeles, 90095, USA.

*Corresponding author. Email: mjarrahi@ucla.edu

**Abstract:** Terahertz waves hold immense potential across diverse fields, including healthcare monitoring, biomedical imaging, precision navigation, high-speed communication, security screening, industrial quality control, and space exploration. However, the widespread adoption of terahertz technology has been hindered by the bulky, complex, and costly nature of existing systems. Here, we demonstrate gain-enhanced interband photomixing in quantum well (QW) PIN photodiodes as an efficient mechanism for frequency-tunable terahertz generation and detection, achieving significant improvements in power efficiency and sensitivity over the state-of-the-art. QWs embedded in PIN photodiodes—key elements of commercially available photonic integrated circuits (PICs)—enable monolithic integration of lasers, semiconductor optical amplifiers (SOAs), modulators, filters, demultiplexers, and other passive optical components. By establishing QW PIN photodiodes as the foundation of a Monolithically Integrated Terahertz Optoelectronic (MITO) platform, this work paves the way for compact, scalable terahertz optoelectronic systems with applications in high-speed data transfer, spectroscopy, and hyperspectral imaging. This advancement positions terahertz technology for widespread use, facilitating practical applications across remote sensing, communications, and medical diagnostics within portable devices.

## Introduction

The terahertz frequency range enables advanced capabilities in sensing, communication, and imaging. Radar systems operating at these frequencies enhance surveillance with smaller apertures, making them ideal for mobile applications such as self-driving cars and security screening. Terahertz carriers also support high-data-rate communication in unallocated frequency bands, while their ability to penetrate opaque materials enables non-destructive quality control. Furthermore, the presence of numerous chemical signatures within this range enables remote identification through spectroscopy. In medical imaging, the low scattering of terahertz waves in biological tissues enables applications such as tumor detection and vascular mapping. However, the strong absorption by water—prevalent in biological systems—should be carefully considered when defining the scope and practical implementation of terahertz medical imaging systems.

While electronic integrated circuits based on field-effect and heterojunction bipolar transistors have been widely used to develop cost-effective and compact terahertz electronic systems[1], monolithically integrated terahertz optoelectronic systems have yet to be realized. This is despite the distinct advantages of optical systems for handling high-frequency signals, including simpler signal distribution and multiplexing, lower propagation losses, and broader modulation and amplification bandwidths[2-11]. Optoelectronic approaches—particularly those relying on photoconductive antennas and nonlinear optical processes—have been developed to implement essential terahertz imaging, sensing, and communication components. However, these systems remain limited in mainstream applications due to their complexity, high cost, and bulkiness. These limitations stem from the dependence on specialized optical sources, amplifiers, modulators, photoconductors, photodiodes, and nonlinear crystals, which are largely incompatible with standard integrated photonics. Additionally, these devices require independent packaging and control electronics, along with multiple optical components to couple the optical pump beam to the device's active area. Without monolithically integrated ultrafast optoelectronic systems, implementing complex imaging, spectroscopy, spectrometry, and communication modalities that rely on arrays of optoelectronic sources and detectors with independent amplitude, phase, and frequency control—such as phased arrays and focal plane arrays—remains impractical. Achieving truly transformative advancements in terahertz technology necessitates a Monolithically Integrated Terahertz Optoelectronics (MITO) platform that seamlessly integrates both active and passive optical components. Such a platform would address the practical challenges of large size, high cost, and system complexity, enabling terahertz optoelectronic technology to transition from laboratory research to widespread real-world applications.

Previous work on waveguide-coupled terahertz sources and detectors has lacked compatibility with photonic integrated circuits (PICs) that incorporate both passive and active optical components[12-20]. Meanwhile, demonstrated integrated terahertz optoelectronics have relied on complex fabrication processes, such as multiple epitaxial growth steps[21,51] to achieve the monolithic integration of uni-traveling-carrier photodiodes with optical pump lasers, semiconductor optical amplifiers (SOAs), and/or modulators, as well as bonding multiple III-V semiconductor stacks onto a silicon photonics chip[22] to heterogeneously integrate lasers, modulators, and photodiodes. Beyond scalability issues, these integrated terahertz optoelectronics schemes have been limited in the operation frequency without demonstrating high-frequency signal detection.

Here, we demonstrate that interband photomixing in quantum well (QW) PIN photodiodes—previously unexplored for terahertz applications—provides an efficient and scalable mechanism for both terahertz generation and detection. By leveraging gain-enhanced photomixing within a platform inherently compatible with photonic integrated circuits (PICs), this approach overcomes key limitations of existing photomixers in efficiency, sensitivity, and integrability. In a prototype fabricated on a GaAs/AlGaAs QW PIC substrate, we demonstrate frequency-tunable terahertz generation and detection across the 100–500 GHz range, achieving both higher terahertz generation efficiency and improved terahertz detection sensitivity compared to the state-of-the-art[23-25]. The compatibility of QW PIN photodiodes with commercially available PIC foundry processes further enables monolithic integration of terahertz sources, detectors, and optical components on a single chip.

## Results

While QW PIN photodiodes are well-established for photonic integrated circuit implementation, their potential for terahertz generation and detection has yet to be realized. Previous studies have been limited to photomixing schemes utilizing intersubband photon absorption in QW structures[26,27], which require pump photons in the long-wavelength regime (mid-infrared and terahertz), where no monolithically integrated platform exists and there are many challenges preventing the integration of essential system components—such as pump sources, detectors, amplifiers, modulators, and passive optical components—onto the same substrate while operating at room temperature. On the other hand, photomixers utilizing interband photon absorption in QW PIN photodiodes have not been previously explored for terahertz generation and detection due to the misconception that the QW energy barrier restricts the ultrafast carrier dynamics necessary for efficient operation. However, our studies on the ultrafast dynamics of the carriers generated through interband absorption in QW structures reveal substantial potential for the development of monolithically integrated terahertz sources and detectors, seamlessly integrated alongside other optoelectronic components.

Under reverse bias, which governs interband photon absorption within the QWs via the quantum-confined Stark effect[28] – the photo-generated carriers escape from the QWs and drift across the intrinsic region (Fig. 1a inset). We model carrier dynamics using a semiclassical quantum-escape formulation: photocarriers generated in the $m^{th}$ QW escape the well by a combination of field-assisted tunneling and thermionic emission (computed using the field-modified density of states and energy-dependent transmission probability) and are then accelerated to saturation velocity across the intrinsic and depleted cladding layers[29]. The resulting induced current is obtained from the carrier motion using the Shockley-Ramo theorem, which accounts for both conduction and displacement currents (see Supplementary Figs. S1). This framework maps material and device parameters (e.g., barrier heights, doping densities, depletion widths, and layer thicknesses) to carrier escape, trapping, and transit times and therefore to a predicted photocurrent. As a result, the contribution of photo-generated electrons and holes in the $m^{th}$ QW traveling across the intrinsic region at high frequencies can be approximated as[29]

$$I_{e,m}(\omega) \approx \frac{z_m}{L+W_n+W_p} \times \frac{1}{1+j\omega\tau_{RC}} \times \frac{qSN_m}{1+j\omega\tau_{QW}^{e}} \times sinc\left(\frac{\omega\tau_{trans}^{e}}{2}\right) e^{-\frac{j\omega\tau_{trans}^{e}}{2}} \quad (1)$$

$$I_{h,m}(\omega) \approx \frac{L-z_m}{L+W_n+W_p} \times \frac{1}{1+j\omega\tau_{RC}} \times \frac{qSN_m}{1+j\omega\tau_{QW}^{h}} \times sinc\left(\frac{\omega\tau_{trans}^{h}}{2}\right) e^{-\frac{j\omega\tau_{trans}^{h}}{2}} \quad (2)$$

where $z_m$ is the distance from the $m^{th}$ well to the N-side, $L$, $W_n$ and $W_p$ are the thicknesses of the intrinsic region, depleted N-region, and depleted P-region, respectively. $S$ denotes the device area, and $N_m$ is the number of photo-generated electrons and holes in the $m^{th}$ QW. $\tau_{QW}^{e}$ and $\tau_{QW}^{h}$ correspond to the electron and hole escape times from the QW respectively, whereas $\tau_{trans}^{e}$ and $\tau_{trans}^{h}$ denote the electron and hole transit times through the intrinsic region. Additionally, $\tau_{RC}$ represents the RC time constant of the device. The total frequency response is obtained by summing the current contributions from all QWs.

The overall frequency response is primarily governed by three key time constants: $\tau_{RC}$ the photodiode's RC time constant; $\tau_{trans}$ the carrier transit time from the QWs to the P/N layers; and $\tau_{QW}$ the carrier escape time from the QWs. The values of $\tau_{QW}$ and $\tau_{trans}$ depend on the reverse bias voltage, as well as the composition and thickness of the heterostructure layers forming the QW PIN photodiode. Meanwhile, $\tau_{RC}$ is dictated by the photodiode's geometry, which determines its capacitance and resistance. Figure 1a presents the theoretically predicted frequency response of a reverse-biased QW PIN photodiode with an RC time constant of 1.55 ps (see Supplementary Figs. S1-S6 for detailed RC time constant and frequency response calculations), fabricated on the GaAs/AlGaAs QW PIC substrate used in our demonstrations. The substrate structure, detailed in Supplementary Fig. S2, provides estimated saturation velocities of $0.72\times10^7$ and $0.8\times10^7$ cm/s for the photo-generated electrons and holes transiting through the depletion region[30,31], along with QW escape times of 0.09 and 0.13 ps for electrons and holes, respectively[29] (see Supplementary Figs. S1 for QW escape time calculations). The contributions of $\tau_{RC}$, $\tau_{trans}$, and $\tau_{QW}$ to the frequency response are depicted in purple, red, and brown, respectively, highlighting the negligible impact of $\tau_{QW}$ on the frequency roll-off compared to $\tau_{RC}$ and $\tau_{trans}$. Comparing this frequency response with that of conventional photomixers, which rely on ultrafast carrier relaxation and recombination across the bandgap, reveals that the carrier transit time in QW PIN photodiodes effectively replaces the role of carrier lifetime in conventional photomixers. The magnitude of this transit time, and therefore the ultimate bandwidth, is strongly influenced by the carrier saturation velocity and effective mass of the semiconductor material, suggesting that alternative material systems may further enhance performance. Thus, careful design of the QW PIN photodiode allows for minimizing $\tau_{RC}$ and $\tau_{trans}$, reducing frequency roll-off and extending the device's operational range within the terahertz spectrum. For instance, reducing the thickness of the intrinsic region while maintaining the same QW structures would decrease the carrier transit time while preserving QW properties for stimulated photon generation, amplification, and phase/intensity modulation. Similarly, reducing the device active area and thickness of the P/N layers as well as increasing their doping would lower the RC time constant while maintaining the QW properties.

Leveraging the ultrafast dynamics of carriers generated through interband photon absorption, an optical pump beam containing two frequency components separated by a terahertz frequency difference can generate a terahertz photocurrent through photomixing (Fig. 1b). The frequency of the resulting terahertz signal can be tuned by adjusting the optical beat frequency, $f_{beat}$, with the generated terahertz power scaling quadratically with the induced photocurrent—up to saturation at high reverse bias voltages and optical pump power levels. Similarly, by coupling a received terahertz signal at $f_{THz}$ to a reverse-biased QW PIN photodiode that is simultaneously pumped with a terahertz beat frequency, an intermediate frequency (IF) photocurrent is induced at $|f_{THz} - f_{beat}|$ through photomixing (Fig. 1c). By tuning the optical beat frequency near the terahertz frequency of interest, the resulting IF signal falls within the radio frequency (RF) range, making it compatible with standard RF electronics for straightforward processing.

As previously mentioned, QW PIN photodiodes serve as the fundamental building blocks of commercially available PICs, enabling the integration of active and passive optical components onto the same substrate. Figures 1b–g illustrate how the same QWs, embedded within the intrinsic region of a PIN photodiode waveguide, can function as multiple optoelectronic components, including a terahertz source (Fig. 1b), a terahertz detector (Fig. 1c), an optical pump source (Fig. 1d), an optical amplifier (Fig. 1e), an intensity modulator (Fig. 1f), and a phase modulator (Fig. 1g). Each function is determined by the operational mode of the device. When current is injected into the photodiode, stimulated photon emission enables it to operate as a laser diode, producing one or multiple emission wavelengths depending on the optical feedback mechanism. Narrowband reflecting facets (e.g., distributed Bragg reflectors) enable single-wavelength emission, while broadband reflecting facets allow multi-wavelength operation[32,33] (Fig. 1d). In the absence of highly reflective facets, the photodiode can function as a semiconductor optical amplifier (SOA), amplifying an incoming optical beam through stimulated photon emission as it propagates through the photodiode waveguide (Fig. 1e). Under reverse bias, the quantum-confined Stark effect[28] allows the photodiode to operate as an intensity modulator (Fig. 1f) and a phase modulator (Fig. 1g) since variations in the QW absorption spectrum provide precise control over both light intensity and phase[34,35]. Furthermore, QW intermixing is a very effective way to obtain low loss waveguides and passive components such as distributed Bragg reflectors, MMI couplers, optical cavities, wavelength demultiplexers and multiplexers, on the same QW PIN substrate[36-38].

Figure 2a illustrates a fabricated terahertz source/detector prototype that generates and detects terahertz signals via gain-enhanced photomixing within a GaAs/AlGaAs QW PIN photodiode (see Methods and Supplementary Figs. S2 and S8). The optical pump beam is delivered through a monolithically integrated SOA ridge waveguide on the same substrate. To minimize $\tau_{RC}$ while sustaining high quantum efficiency, a tapered photomixer geometry is employed, which reduces parasitic capacitance. Additionally, the P-cladding is partially etched to keep parasitic resistance low. A tapered transition region between the SOA and photomixer mitigates mode mismatch, ensuring efficient light coupling. The photomixer contacts are connected to ground-signal-ground (GSG) pads to detect/apply terahertz signal through standard GSG probes when operating as a terahertz source/detector. The first prototype features a 12-µm-long photomixer with a tapered waveguide width varying from 3 µm to 0.5 µm terminated with an output probe impedance of 50 Ω. This configuration is estimated to offer a $\tau_{RC}$ of 1.55 ps and 56% optical absorption in the photomixer active region at a 3 V reverse bias (see Supplementary Figs. S4 and S5). Figure 2b shows the optical mode transition from the SOA to the photomixer, demonstrating how the tapered design focuses light into the photomixer while preserving high quantum efficiency. To electrically isolate the SOA from the photomixer, the tapered transition region is partially proton-implanted, providing a resistance of 1 GΩ for a 5 µm-long implanted region. Figure 2c illustrates the transmitted optical power through varying lengths of the implanted region, indicating an optical loss of 0.21 dB/µm.

Two commercially available lasers are used to characterize both the SOA and gain-enhanced photomixer that form the terahertz source and detector. The two lasers can be replaced by two wavelength tunable lasers fabricated on the same QW PIN photodiode substrate[39]. Figure 2d shows the output power of a 1-mm-long SOA as a function of the SOA pump current, with an input optical power of 0.15 mW, indicating a threshold current of approximately 60 mA, and a free-space-coupled maximum output power of 8 mW at a pump current of 130 mA (see Methods). The relative intensity of the SOA input and output spectra for an optical beat frequency of 300 GHz is shown in the insets of Fig. 2d. The slight imbalance between the two amplified tones is attributed to the wavelength dependence of the SOA gain, as evidenced by the amplified spontaneous emission (ASE) spectrum.

Figure 3a shows the generated power from the terahertz source/detector as a function of the photomixer photocurrent at 230 GHz, with an input optical power of 5 mW (see Methods). The photomixer photocurrent can be tuned by both the SOA pump current (i.e., the optical power pumping the photomixer) and the photomixer bias voltage, as shown in the inset of Fig. 3a. While the photomixer photocurrent remains relatively constant for bias voltages between -1.5 V and -3 V (suggesting that the external quantum efficiency reaches its maximum at -1.5 V), the measured terahertz power at the same photocurrent level increases significantly when the bias voltage is increased from -1.5 V to -3 V. This increase is attributed to the reduction in the photo-generated electron/hole energy barrier, which substantially shortens their escape time from the QWs. In the absence of frequency roll-off, the generated signal power is expected to follow the theoretical expression of $\frac{1}{2} I_{ph}^2 \times 50\Omega$ (marked by the dashed black line), where $I_{ph}$ is the photomixer photocurrent and 50 Ω is the impedance of the GSG probe. While the measured signal powers adhere to this theoretical slope up to the saturation point at higher photocurrent levels, the observed deviations from the theoretically predicted powers are due to frequency roll-off, which is caused by device parasitics, carrier transit time, and carrier escape time from the QWs. Figure 3b shows the maximum generated power from the terahertz source/detector as a function of frequency at a bias voltage of -3 V. The SOA gain is estimated to be 6.2 dB at 5 mW input optical power (see Methods). The 3 MHz 3-dB linewidth of the generated terahertz signal, shown in the inset of Fig. 3b, and its phase noise (see Supplementary Fig. S9) are directly determined by the linewidth of the two free-running lasers that produce the optical beat signal for these measurements.

Figure 3c shows the conversion loss of the terahertz source/detector when down-converting a 240 GHz signal to 0.8 GHz, with an input optical power of 5 mW (see Methods). Similar to the terahertz generation measurements, the photomixer photocurrent is varied by adjusting the bias voltage and the SOA pump current. The optimal operating conditions for high-sensitivity terahertz detection are determined by considering the tradeoff between the conversion loss and noise performance of the photomixer. The noise is dominated by Johnson-Nyquist noise, shot noise from the photomixer, and amplified spontaneous emission (ASE) from the SOA. As shown in the inset of Fig. 3c, the output noise power of the photomixer exhibits a linear dependence on the photocurrent. The highest signal-to-noise ratio for terahertz detection is achieved at a photomixer bias voltage of -0.7 V and a photocurrent of 0.38 mA. These conditions are therefore used for the subsequent terahertz detection measurements. Figure 3d displays the measured conversion loss and terahertz detection sensitivity (specified as the input-referred noise power density) as a function of frequency under these operating settings. The spectrum of the down-converted signal is processed using a real-time spectral lock-in detection code (see Methods and Supplementary Fig. S10), which offers an input-referred noise power density reduction rate of 10 dB per decade as a function of integration time while preserving spectral information (Fig. 3d inset).

By leveraging interband photomixing, ultrafast carrier dynamics, the quantum-confined Stark effect, and stimulated photon emission/amplification provided by the QW structures, the demonstrated gain-enhanced QW photomixer prototype achieves frequency-tunable terahertz generation and detection across the 100–500 GHz range. Notably, it outperforms photomixer-based terahertz sources in terms of terahertz generation efficiency (Fig. 4a) while also surpassing photomixer-based terahertz detectors in detection sensitivity (Fig. 4b). Two key factors contribute to this performance enhancement. First, the relatively long interaction length between the optical pump beam and the QW photo-absorbing region (when compared with surface-illuminated photomixers), where the photoabsorbed carriers can escape from and transit to the P/N layers in a sub-picosecond timescale, enabling both high quantum efficiency and ultrafast photomixing. Second, the gain-enhanced photomixing process within a monolithically integrated platform, enhancing the terahertz generation/detection efficiency. While the QW PIN photomixer can generate/detect terahertz signals via interband photomixing when directly pumped by two lasers with a terahertz beat frequency, integrating an SOA significantly reduces the required optical power and enhances the terahertz generation/detection efficiency. Our estimates indicate that the SOA gain increases the terahertz generation efficiency of the demonstrated sources by an order of magnitude (see Supplementary Fig. S11). This highlights a key advantage of QW PIN photodiodes enabling scalable, complex terahertz optoelectronic systems based on large source/detector arrays.

Whereas two optical tones with a terahertz frequency difference generate and detect a single-tone terahertz signal, a multi-tone optical pump beam enables multi-tone terahertz generation and detection through interband photomixing (see supplementary Fig. S12). Figure 5a illustrates multi-tone terahertz generation using a Fabry-Pérot laser fabricated on the same GaAs/AlGaAs QW PIC substrate. The Fabry-Pérot laser waveguide is cleaved, and its multi-tone optical emission—with a total power of 10 mW—is edge-coupled to the SOA-integrated photomixer, where the optical tones are amplified and mixed. The resulting multi-tone signal, generated through photomixing, contains the beat frequencies of all possible optical tone pairs (see Supplementary Fig. S13).

Beyond pump lasers, which may be multi-tone, single-tone[39, 42-45], frequency comb[46], or mode-locked[47,48] – enabling both frequency-domain and time-domain terahertz spectroscopy and imaging – other optoelectronic system components can be fabricated on the same QW substrate without requiring epitaxial regrowth or chip bonding processes. Figures 5b and 5c present the intensity and phase modulation performance of a QW PIN waveguide fabricated on the same GaAs/AlGaAs PIC substrate. Under an applied reverse bias voltage, the QW energy levels shift, reducing the bandgap energy due to the quantum-confined Stark effect[28]. This results in a redshift of the absorption spectrum (Fig. 5b inset) and modifies both the absorption coefficient and refractive index of the waveguide. Consequently, this enables intensity modulation with an extinction ratio of 21 dB for a 100-μm-long waveguide and phase modulation with a $V\pi.L$ of 0.05 V·mm. As illustrated in Figs. 5b and 5c, the quantum-confined Stark effect induces both intensity and phase modulation—an intrinsic feature of any modulation process in linear optical systems. According to the Kramers–Kronig relations, and as demonstrated across various PIC platforms (including III-V quantum wells, lithium niobate, and silicon photonics), intensity and phase modulation are inherently coupled in linear regimes. Consequently, the tradeoff between the intensity and phase modulation must be carefully considered in the design of MITO systems. The ability of QW PIN PIC platforms to provide optical gain offers a key advantage, helping to mitigate this tradeoff and maintain high system performance.

**Discussion**

These examples represent just a fraction of the many optoelectronic functionalities that can be obtained on the same QW PIN substrate, which, as demonstrated in this work, supports both high-efficiency terahertz generation and high-sensitivity terahertz detection. In fact, high-performance tunable lasers, SOAs, modulators, filters, demultiplexers, and other passive optical components monolithically integrated on QW PIN substrates are already available through commercial PIC foundry processes[49]. This establishes QW PIN photodiodes as the core building block of a MITO platform, making it possible to realize chip-scale optoelectronic terahertz imaging, spectroscopy, and communication systems for the first time. For instance, Figure 6 illustrates how the introduced MITO platform can enable the development of an optoelectronic terahertz phased-array transceiver for adaptive hyperspectral remote sensing and communication on a chip.

While QW PIN structures possess the essential physical characteristics to support a wide range of functionalities—including terahertz sources and detectors via gain-enhanced interband photomixing, lasers and SOAs through stimulated photon emission, modulators leveraging the quantum-confined Stark effect, and passive components such as couplers, filters, and demultiplexers enabled by QW intermixing—it is challenging to design a single structure that offers optimal performance for all these applications. For instance, reducing the thickness of the QWs can improve photodetection speed by reducing carrier transit time, but at the expense of reduced modulation depth. Likewise, increasing P and N doping concentrations can enhance carrier injection for lasers and SOAs and improve photodetector response time, while simultaneously increasing free carrier absorption in passive devices. These tradeoffs are well recognized, and commercial III-V foundries have developed QW PIN structures that offer a balanced compromise across different functionalities. In our demonstration, we used a commercially available GaAs/AlGaAs QW PIC substrate to realistically account for these tradeoffs, even though it was not specifically optimized for terahertz generation and detection. Nonetheless, our results highlight the strong potential of QW PIN structures for high-performance terahertz generation and detection on a monolithically integrated platform, and suggest that further optimization of the QW design could yield even greater performance improvements.

While the present demonstration is based on a GaAs/AlGaAs quantum well platform, the underlying gain-enhanced interband photomixing mechanism is broadly applicable to other semiconductor material systems. The achievable terahertz bandwidth and efficiency are primarily governed by carrier escape time, carrier transit velocity, and device parasitics, which in turn depend on the effective mass, band offsets, and saturation velocity of the semiconductor. Materials with higher carrier saturation velocities and lower effective masses, such as InGaAs/InAlAs or InP-based QW structures, could enable faster carrier transport and extended terahertz bandwidth. Emerging material systems, including GaN/AlGaN, may further extend the operational range toward higher frequencies due to their large band offsets and potentially ultrafast carrier dynamics. These considerations highlight that the proposed MITO platform is not limited to a specific material system but can be adapted to different semiconductor platforms depending on the targeted performance metrics and integration requirements.

Beyond the terahertz source and detector architecture demonstrated in this work, the monolithic integration of photomixers with lasers, SOAs, and other optical components opens the door to higher-performance optoelectronic terahertz sources and detectors based on distributed arrays and traveling-wave architectures. Just as integrated circuit technology transformed early, bulky, power-hungry computers into the high-performance microprocessor chips in use today in homes, cars, phones, and health monitoring systems, a MITO platform based on QW PIN terahertz sources and detectors could bring about a similar technological transformation in the terahertz field. It could enable a transition from expensive, bulky, laboratory-grade optoelectronic terahertz imaging, sensing, spectrometry, and communication systems to scalable, compact, and low-cost microchips suitable for widespread, real-world applications.

## Methods

### Fabrication Process

The fabrication process, depicted in Supplementary Fig. S8, begins with lithographic patterning of the SOA top contacts, followed by deposition of a 10/300 nm Cr/Au layer and a liftoff step. The SOA ridge waveguides are then patterned and formed by etching down to the etch-stop layer, utilizing a combination of reactive ion etching and wet etching (see Supplementary Fig. S8a). Following this, the photomixer top contacts are patterned via e-beam lithography, with 10/400 nm Cr/Au deposited and subsequently lifted off. The ion-implantation regions are then defined using photolithography, after which the sample undergoes proton implantation at room temperature, using a dosage of $5\times10^{14}$ $cm^{-2}$ at 70 keV and a 7° tilt angle (see Supplementary Fig. S8b). Next, tapered transition regions are defined lithographically and formed by reactive ion etching of a 200-nm-thick AlGaAs layer (see Supplementary Fig. S8c). The SOA and shallow taper regions are then protected with photoresist, and the photomixer top contacts serve as a hard mask for forming the photomixer waveguides via dry etching (see Supplementary Fig. S8d). Subsequently, AlGaAs regions outside the device core are removed through a combination of dry and wet etching to reach the underlying n+ GaAs layer. The SOA and photomixer bottom contacts are patterned using photolithography, followed by deposition of a 75/300 nm AuGe/Au layer, liftoff, and rapid thermal annealing at 380 °C for 30 s. An additional 2 μm of Au is then deposited on the photomixer bottom contacts to raise them to the same height as the top contacts, allowing for GSG probe placement (see Supplementary Fig. S8e). Benzocyclobutene (BCB) is subsequently spin-coated, cured, and etched back to planarize the device surface, followed by additional metal deposition in the SOA regions to form the top contact pads (see Supplementary Fig. S8f). BCB is then fully removed from the GSG probe pad areas, along with etching of the underlying n+ GaAs layer (see

Supplementary Fig. S8f). Another round of BCB spin coating and etch-back follows, exposing both the SOA and photomixer contacts (see Supplementary Fig. S8g). Finally, a 50/450 nm Ti/Au layer is deposited to form the GSG probe pads (see Supplementary Fig. S8h), and the sample is cleaved in preparation for testing.

**SOA Characterization Setup**

The fabricated SOA samples are cleaved to form 1-mm-long straight ridge waveguides. These SOA waveguides are oriented at a 7° angle relative to the [100] direction of the wafer. To reduce reflections at a wavelength of approximately 809 nm, a 104-nm-thick $Al_2O_3$ anti-reflection coating is evaporated on both facets of the SOA samples. For initial testing, an optical signal from a DBR laser operating at ~809 nm (Thorlabs DBR808PN) is coupled into the SOA via a lensed fiber. The output beam is then collected using a 4f imaging system and measured with an optical power meter, as shown in Supplementary Fig. S14a. The input optical power coupled into the SOA is estimated based on the measured photocurrent of the SOA under reverse bias. To characterize the spectral properties of the SOA, two optical beams from DBR lasers (both operating at ~809 nm; Thorlabs DBR808PN) are combined through a 50:50 fiber coupler and coupled into the SOA using a lensed fiber. The SOA output beam is directed to an optical spectrum analyzer (OSA) via a lens and fiber collimator, as illustrated in Supplementary Fig. S14b.

**Terahertz Source Characterization Setup**

The block diagram of the terahertz source characterization setup is shown in Supplementary Fig. S15a. The optical beams from two wavelength-tunable DBR lasers operating at ~809 nm (Thorlabs DBR808PN) with a terahertz beat frequency are combined using a 50:50 fiber coupler and coupled into the SOA through a lensed fiber. Ground-signal-ground (GSG) terahertz probes, covering frequency bands of 140–220 GHz, 220–330 GHz, and 330–500 GHz (FormFactor T-Wave probes), are employed to route the generated terahertz signal to a harmonic mixer (VDI SAX) for down-conversion to an intermediate frequency (IF) around 1.2 GHz. The low-frequency port of the bias-T integrated with the GSG probes is used to apply the photomixer bias voltage while simultaneously recording its DC photocurrent. The IF signal is then amplified by an RF amplifier (Mini-Circuits ZRL-1150LN+) and split into two paths with an RF splitter (Mini-Circuits ZN2PD1-222-S+), allowing for simultaneous monitoring of the IF spectrum using an electrical spectrum analyzer and measurement of IF power with a calibrated RF power meter (HP 438A power meter with HP 8481A power sensor). To reduce out-of-band noise, a bandpass filter (Pasternack PE8731) is placed before the power detector. The conversion loss of the VDI SAX module is calibrated separately for each frequency band. For operation in the 330–500 GHz band, a WR2.2 active multiplier chain (AMC) (VDI WR9.0SGX-M + WR4.3×2 + WR2.2×2) serves as the source, and its power is measured across the spectral range using a calibrated power meter (VDI PM5B) through a WR2.2 to WR10 taper and a 1-inch WR10 waveguide. The measured power values are corrected to account for taper and waveguide losses based on vendor calibration data. The WR2.2 AMC is then connected to the WR2.2 SAX module, followed by the IF electronics, and IF power is measured with the calibrated RF power meter. The total conversion loss of the setup is determined by comparing the measured IF and terahertz power levels, incorporating the GSG probe losses from vendor calibration data. For operation in the 140–220 GHz band, an Eravant SFA-114174402-06VF-E1 AMC is used as the source, and power is measured over the spectral range using a calibrated power meter (VDI PM5B) through a WR5.1 to WR10 taper and a 1-inch WR10 waveguide. In the 220–330 GHz band, a Millitech AMC-10-RFHB0 AMC connected to a VDI WR10×3 frequency tripler is used as the source, with power measured over the entire spectral range using the same calibrated power meter (VDI PM5B) through a WR3.4 to WR10 taper and a 1-inch WR10 waveguide. The remaining calibration process for these bands follows the procedure described for the 330–500 GHz band.

**SOA Gain Estimate**

The SOA gain at an input optical power of 5 mW is estimated from the measured photocurrent of the photomixer. As illustrated in Fig. 3a, the highest terahertz power is achieved at a photomixer bias voltage of -3 V and a photocurrent of $I_{ph}$ = 6 mA. The estimated input optical power to the photomixer is calculated as $\frac{1}{\eta}\frac{hc}{q\lambda}I_{ph} = 16.3$ mW, where $h$ is the Planck's constant, $c$ is the speed of light, $q$ is the electron charge, $\lambda$ is the optical wavelength of 809 nm, and $\eta$ is the photomixer quantum efficiency, estimated to be 56.5% from the electromagnetic simulations (Fig. 2b). The SOA output optical beam propagates through a 5-μm-long implanted electrical isolation region with an optical loss of 0.21 dB/μm (Fig. 2c) before reaching the photomixer, resulting in an estimated SOA output power of 20.76 mW and SOA gain of 6.2 dB.

**Terahertz Detector Characterization Setup**

The block diagram of the terahertz detector characterization setup is shown in Supplementary Fig. S15b. The optical beams from two wavelength-tunable DBR lasers operating at ~809 nm (Thorlabs DBR808PN), with a terahertz beat frequency, are combined via a 50:50 fiber coupler and coupled into the SOA through a lensed fiber. Ground-signal-ground (GSG) terahertz probes, covering the 140–220 GHz, 220–330 GHz, and 330–500 GHz frequency bands (FormFactor T-Wave probes), are used to apply the terahertz signal generated by the sources described in the previous section for each frequency band. The terahertz power at each frequency is measured using a calibrated power meter (VDI PM5B). The down-converted IF signal, centered around 0.8 GHz, is extracted through the low-frequency port of the bias-T integrated with the GSG probe. The DC and RF components of the IF signal are separated using an RF bias-T (Picosecond Pulse Labs 5541A-104). The RF component is amplified by an RF amplifier (Mini-Circuits ZRL-1150LN+), filtered with a bandpass filter (VBFZ-780-S+), and measured by a calibrated RF power meter (HP 438A power meter with HP 8481A power sensor). Simultaneously, the IF spectrum is monitored with an electrical spectrum analyzer using an RF splitter (Mini-Circuits ZN2PD1-222-S+). The conversion loss of the device is calculated by comparing the terahertz power with the IF power, factoring in the gain of the RF electronics. The calibration process follows the same method as that used for terahertz source characterization. To perform spectral lock-in detection, the input terahertz signal is first modulated, and the electrical spectrum analyzer is replaced with a real-time lock-in detection analyzer. This analyzer down-converts the IF spectrum from ~0.8 GHz to ~40 MHz using an RF mixer (RFCOMP HD26217) and then digitizes it with an analog-to-digital converter (Texas Instruments ADS5485). A Fast Fourier Transform is then applied to the time-domain signal, decomposing it into distinct frequency components spanning 0–100 MHz. All frequency components are demodulated in parallel to generate the lock-in detected spectrum[50].

**Data Availability:** All the data and methods needed to evaluate the conclusions of this work are present in the main text and the Supplementary Materials. Additional data can be requested from the corresponding author.

**Acknowledgments:** The authors gratefully acknowledge the financial support from the Office of Naval Research (grant # N000142212531) and IET Harvey Engineering Research Prize. Yifan Zhao was supported by the Department of Energy (grant # DE-SC0016925).

**Author contributions:** Y.Z. designed, fabricated, and characterized the terahertz optoelectronic devices; S.Z. helped with the design, modeling and measurements, S.T. helped with the measurements and data processing, M.J. initiated and supervised the research; all authors contributed to the preparation of the manuscript.

**Competing interests:** M.J. is a co-founder of Lookin Inc. The remaining authors declare no competing interests.

**Correspondence and requests for materials** should be addressed to Mona Jarrahi.

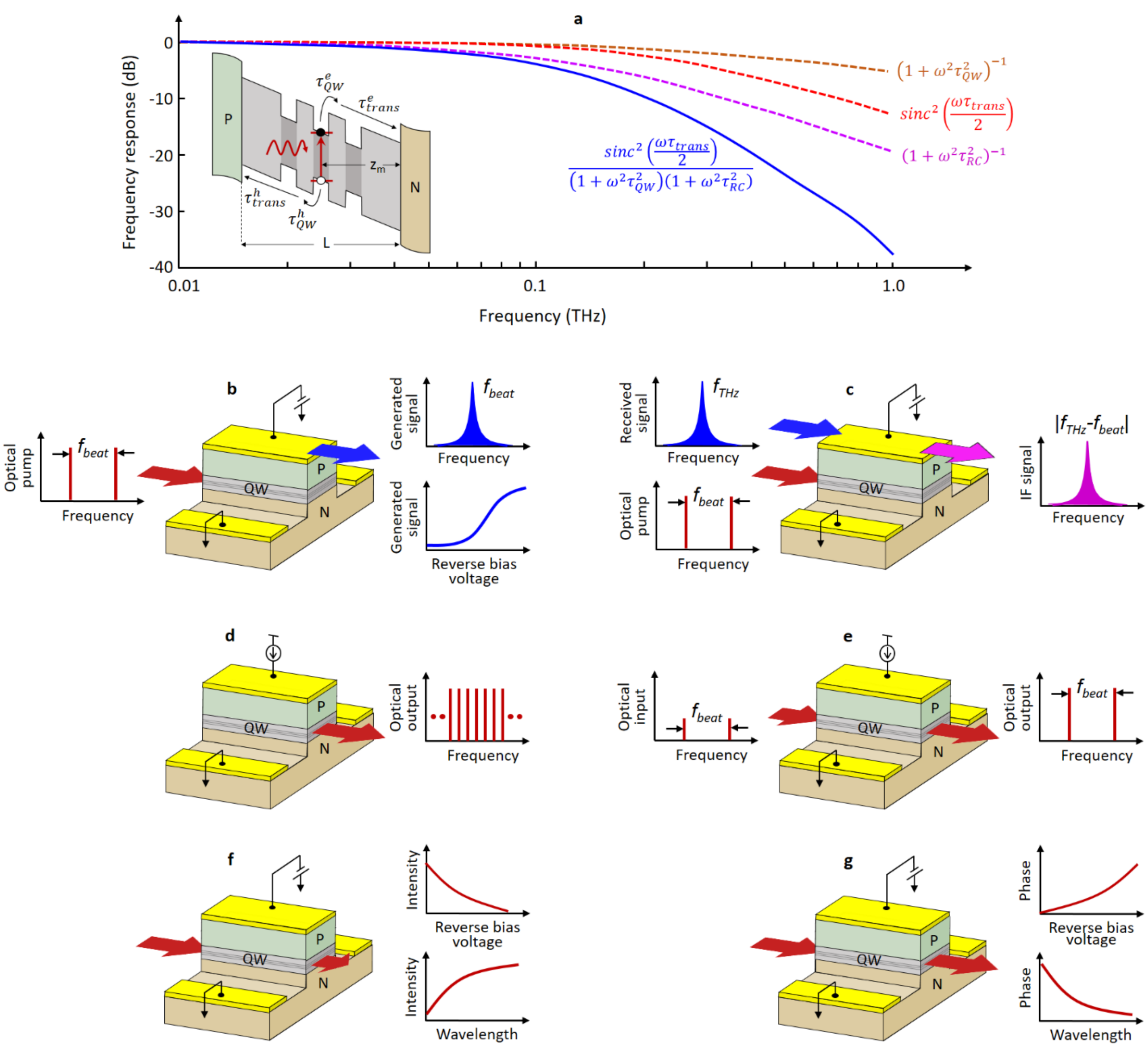

**Fig. 1 | The monolithically integrated terahertz optoelectronics (MITO) platform based on a QW PIN photodiode structure.** **a.** The theoretically predicted frequency response of the interband photomixing process in the QW PIN photodiode shows the contributions of $\tau_{RC}$, $\tau_{trans}$, and $\tau_{QW}$ to the overall response. These predictions assume electron/hole saturation velocities of $0.72\times10^7$ / $0.8\times10^7$ cm/s through the depletion region, QW electron/hole escape times of 0.09/0.13 ps[29], and a photomixer RC time constant of 1.55 ps. The ultrafast carrier dynamics of the interband photomixing process enables the realization of terahertz sources and detectors on the same QW PIN substrate, as illustrated in **b** and **c**, respectively. Other key building blocks of the MITO platform that can be monolithically integrated on the same QW PIN substrate include: **d**, optical pump source, **e**, semiconductor optical amplifier, **f**, intensity modulator, and **g**, phase modulator.

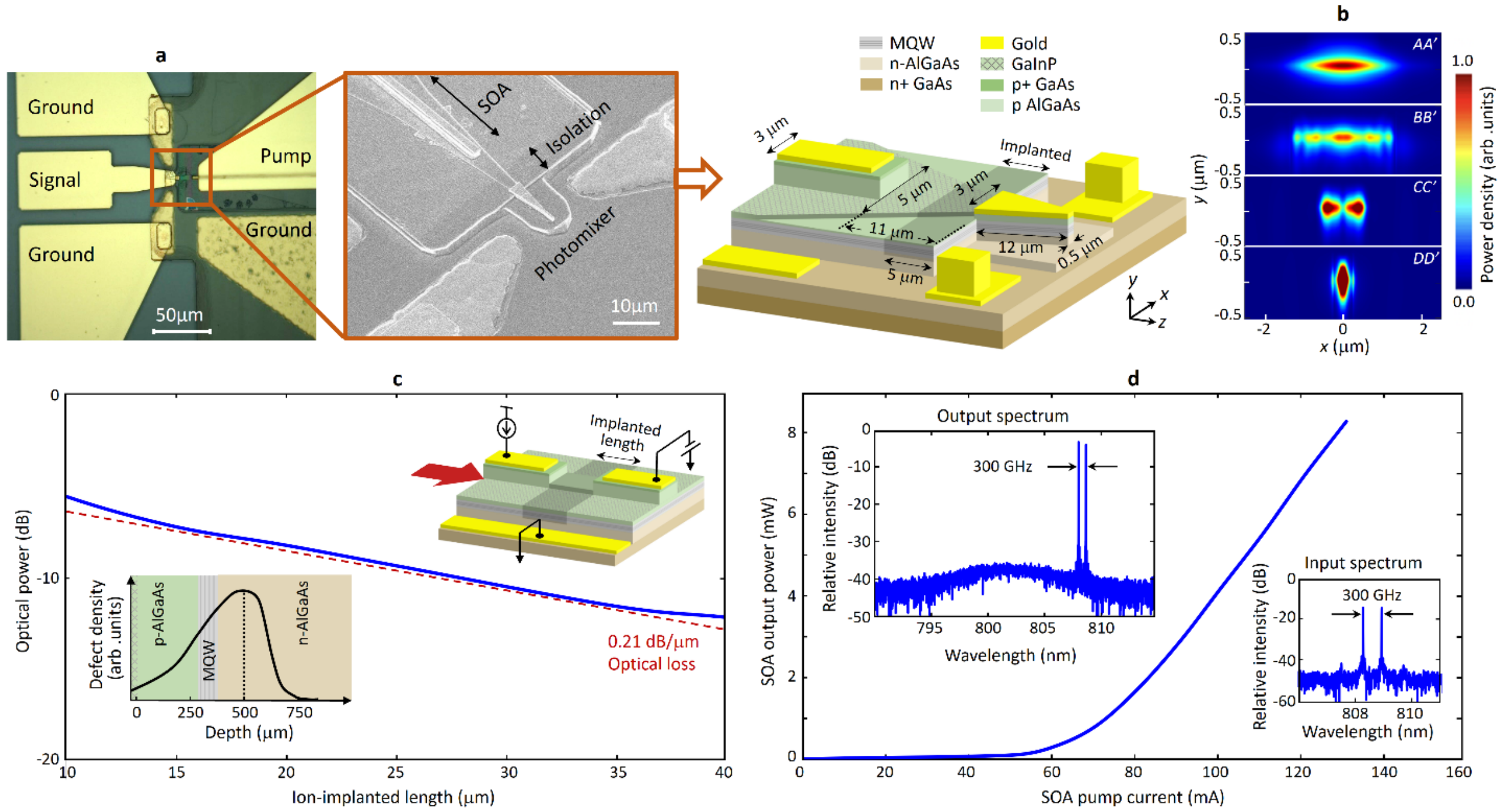

**Fig. 2 | Gain-enhanced photomixer based on a QW PIN photodiode structure. a**, A fabricated terahertz source/detector prototype that generates/detects terahertz signals through gain-enhanced photomixing in a GaAs/AlGaAs QW PIN photodiode. The photomixer waveguide has a tapered width ranging from 3 μm to 0.5 μm over a 12 μm length and is pumped by a monolithically integrated SOA ridge waveguide with a 3 μm ridge width. **b**, The optical mode inside the SOA ridge waveguide (cross section *AA'*) and at distances of 12/17/20 μm from the SOA output facet (cross sections *BB'*/*CC'*/*DD'*). Using electromagnetic simulations in Lumerical, we estimate an optical transmission of 97% through the tapered transition region and an optical absorption of 56.5% within the photomixer's active region. **c**, To characterize the optical loss of the implanted electrical isolation region, we fabricated implanted test structures (top inset) with varying lengths and measured the transmitted optical power through them. The optical beam coupled to each implanted test structure is first amplified via propagation through a forward-biased PIN waveguide. The same waveguide is reverse biased during optical alignment to verify uniform coupling across all test structures using the measured photocurrent. The transmitted optical power through each implanted test structure is detected using a reverse-biased PIN waveguide. The bottom inset shows the simulated lattice defect density using SRIM software to ensure the implantation covers the entire P-region. **d**, SOA output power as a function of pump current for an input optical power of 0.15 mW at an 809 nm wavelength. Insets show the relative intensity of the SOA input and output spectra for an optical beat frequency of 300 GHz.

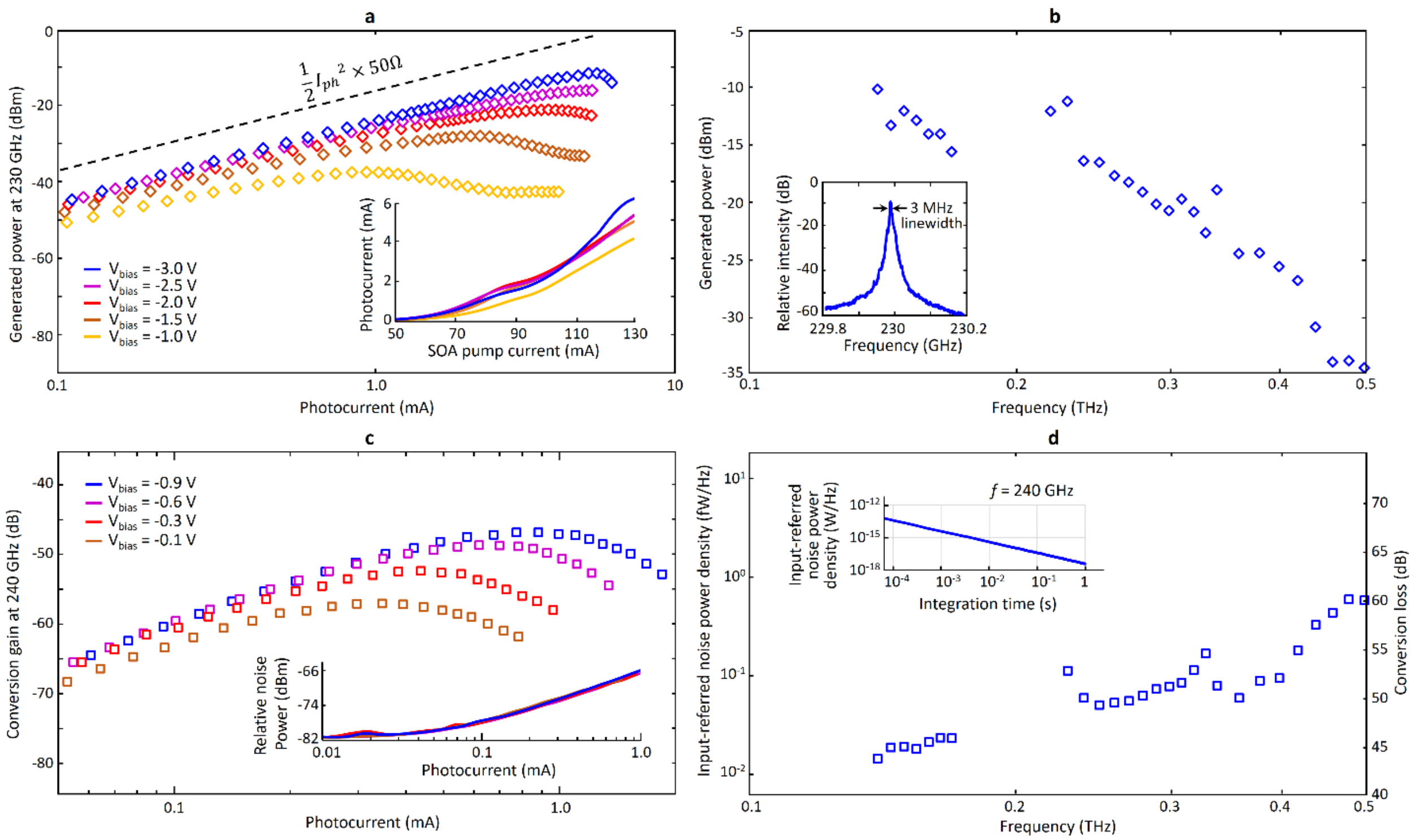

**Fig. 3 | Demonstration of integrated terahertz sources and detectors based on a QW PIN photodiode structure. a**, Generated power from the terahertz source/detector as a function of the photomixer photocurrent at 230 GHz. The inset shows the dependence of the photomixer photocurrent on the SOA pump current and the photomixer bias voltage. **b**, Maximum generated power as a function of frequency at a bias voltage of -3 V. The SOA gain is estimated to be 6.2 dB (see Methods). Power levels are measured using three different harmonic mixers and probes covering the frequency ranges 140-170 GHz, 230-330 GHz, and 340-500 GHz (see Methods). The abrupt power changes at the boundaries of these frequency ranges are due to deviations in the scattering parameters of the GSG probes, waveguide connections, and harmonic mixers used in the measurements. The inset shows the spectrum of the generated signal at 230 GHz. **c**, Conversion loss of the terahertz source/detector as a function of the photomixer photocurrent at 240 GHz. The inset shows the dependence of the photomixer noise power on the photocurrent, observed on an electrical spectrum analyzer. **d**, Conversion loss and input-referred noise power density at a bias voltage of -0.7 V, photocurrent of 0.38 mA, and an integration time of 64 ms. Spectrum of the down-converted signal is processed using a real-time spectral lock-in detection code (see Methods and Supplementary Fig. S10), which offers an input-referred noise power density reduction rate of 10 dB per decade as a function of integration time while preserving spectral information. Inset shows the dependence of the input-referred noise power density on the integration time at 240 GHz for a 15 kHz modulation frequency.

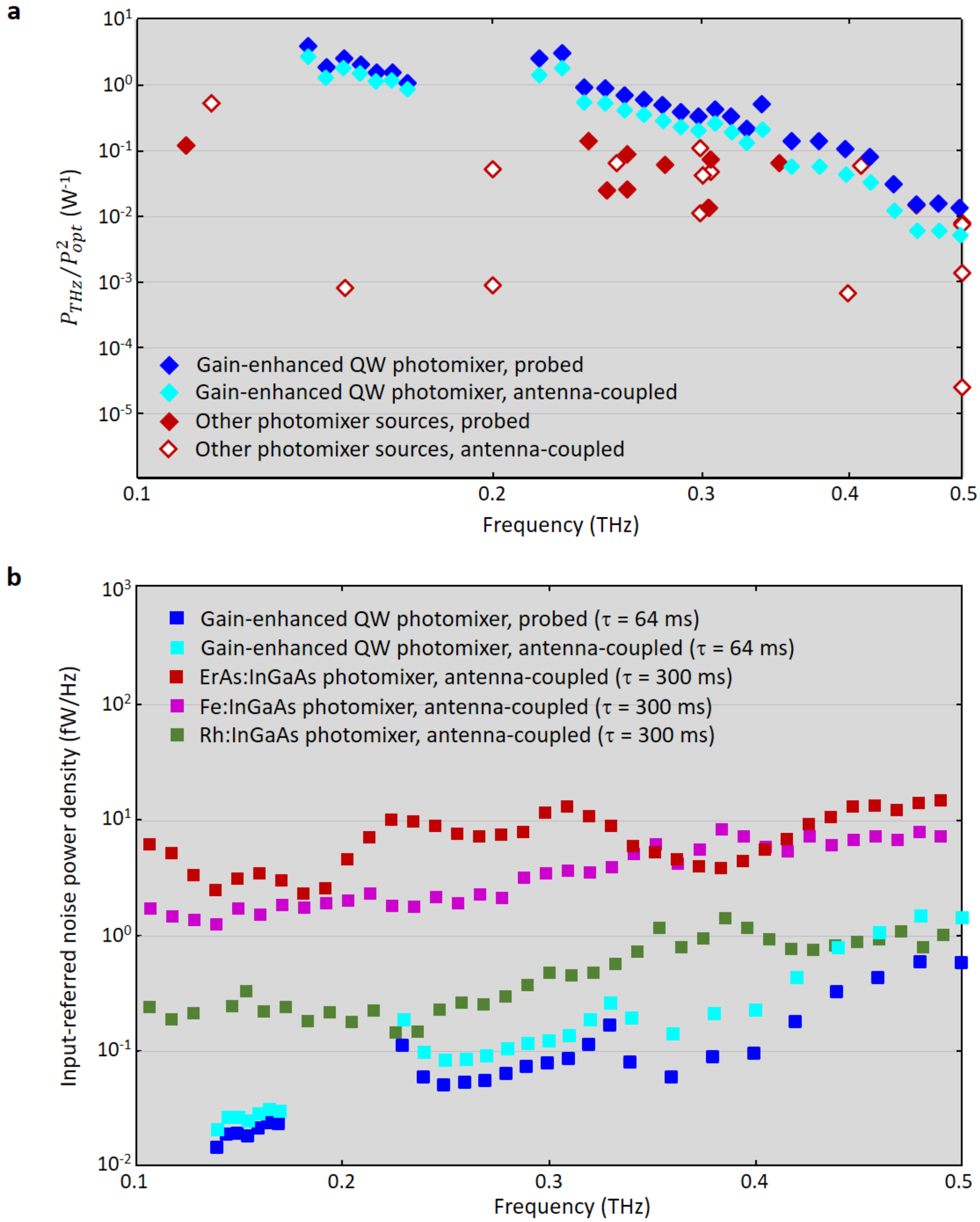

**Fig. 4 | Performance comparison with the state-of-the-art. a**, Comparison of the efficiency figure of merit $P_{THz}/P_{opt}^2$ for the demonstrated gain-enhanced QW photomixer in terahertz generation mode with other photomixers from the literature (see Supplementary Table S1). The probed and antenna-coupled sources are represented by solid and open red diamonds, respectively. The GSG probes enable the conversion of the gain-enhanced QW photomixer into an antenna-coupled terahertz source. They

extract the on-chip terahertz signal and launch it as an electromagnetic wave propagating through the rectangular metallic waveguide output of the probe, which can then be connected to a horn antenna to generate free-space terahertz radiation. By measuring the terahertz power levels at the probe waveguide output and using the specified radiation efficiency of commercial horn antennas, we projected the generated terahertz power levels and efficiencies for this antenna-coupled configuration, shown as cyan diamonds. **b**, Comparison of the input-referred noise power density (noise-equivalent power) of the demonstrated gain-enhanced QW photomixer in terahertz detection mode at a 64 ms integration time with other photomixers from the literature at a 300 ms integration time, including those based on ErAs:InGaAs, Fe:InGaAs, and Rh:InGaAs photomixers[40,41]. The GSG probes also enable operation of the gain-enhanced QW photomixer as an antenna-coupled terahertz detector. In this configuration, incident terahertz radiation is collected by a horn antenna and coupled into the rectangular metallic waveguide input of the GSG probe. The probe converts the guided terahertz wave into an electrical signal delivered to the GSG pads of the on-chip device. The projected input-referred noise power density values for this antenna-coupled scheme are represented by cyan squares.

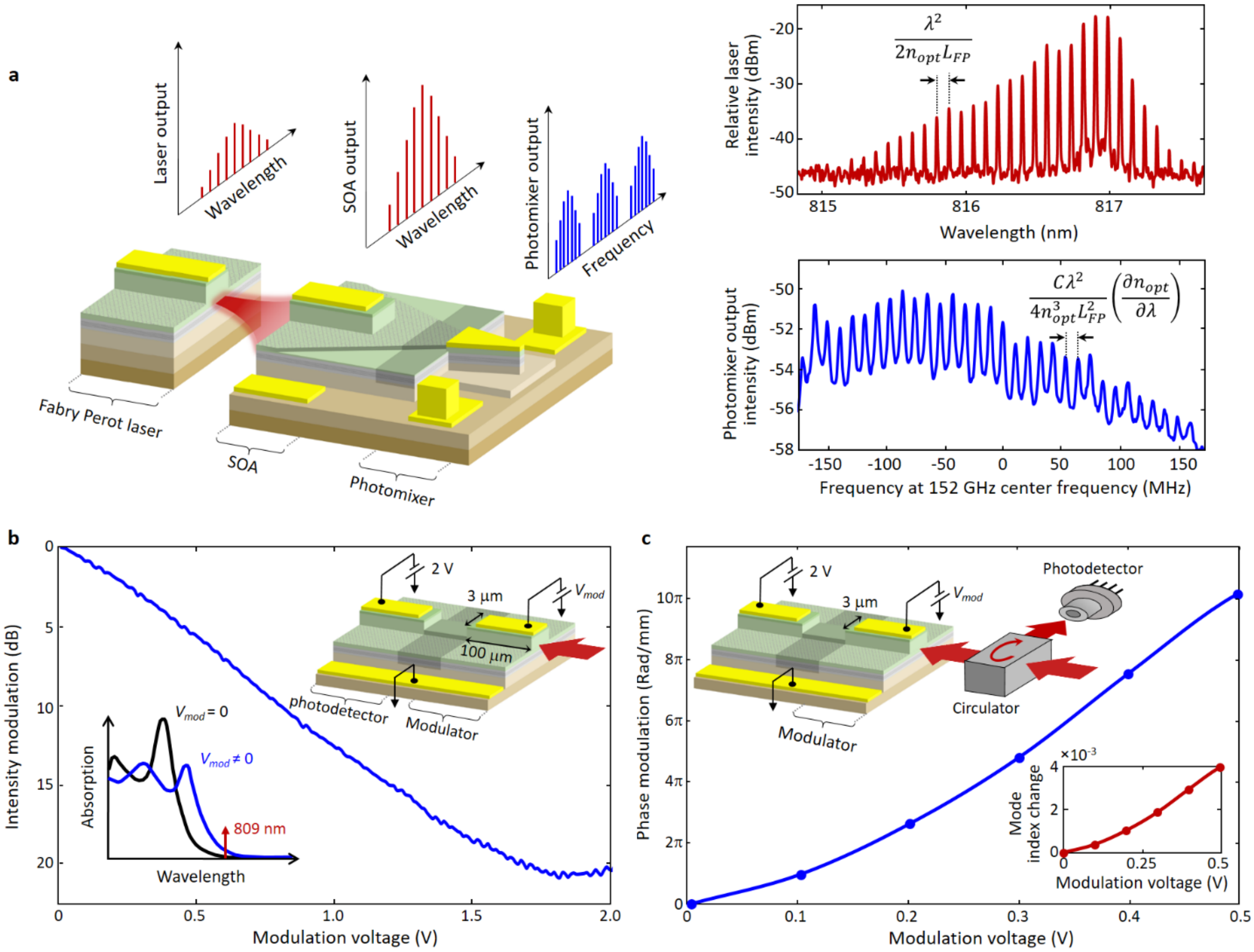

**Fig. 5 | Demonstration of different optoelectronic functionalities offered by the same QW PIN photodiode structure. a**, The multi-tone optical pump beam from a Fabry-Perot laser, fabricated on the same QW PIN substrate, is edge-coupled to the SOA-integrated photomixer, where the optical tones are amplified and mixed. The measured multi-tone laser output for a Fabry-Perot cavity length of $L_{FP}$ = 1 mm (top right ) and the generated multi-tone terahertz signal around 152 GHz (bottom right) are shown. **b,** Optical intensity modulation performance of a 100-µm-long QW PIN waveguide, measured by recording the transmitted optical beam through the QW PIN waveguide as a function of the applied reverse bias voltage (top right inset). The bottom left inset illustrates how the quantum-confined Stark shift in the QW absorption spectrum increases optical absorption at the operating wavelength. **c**, Optical phase modulation performance of the QW PIN waveguide, determined by measuring changes in the optical mode index as a function of the reverse bias voltage (bottom right inset). The top left inset depicts the experimental setup used for measuring the mode index changes. The input optical beam is first directed through a circulator and coupled into the waveguide using a lensed fiber. The optical beam reflected from the waveguide facet is then routed by the circulator to a photodetector. By modulating the reverse bias voltage applied to the waveguide and lock-in detection of the photodetector signal, the corresponding changes in the reflection coefficient of the waveguide mode are determined for each bias voltage. The optical mode index variations,

$\Delta n$, are extracted from these changes in the reflection coefficient as a function of reverse bias voltage. Phase modulation per unit length is calculated as $2\pi\Delta n/\lambda$.

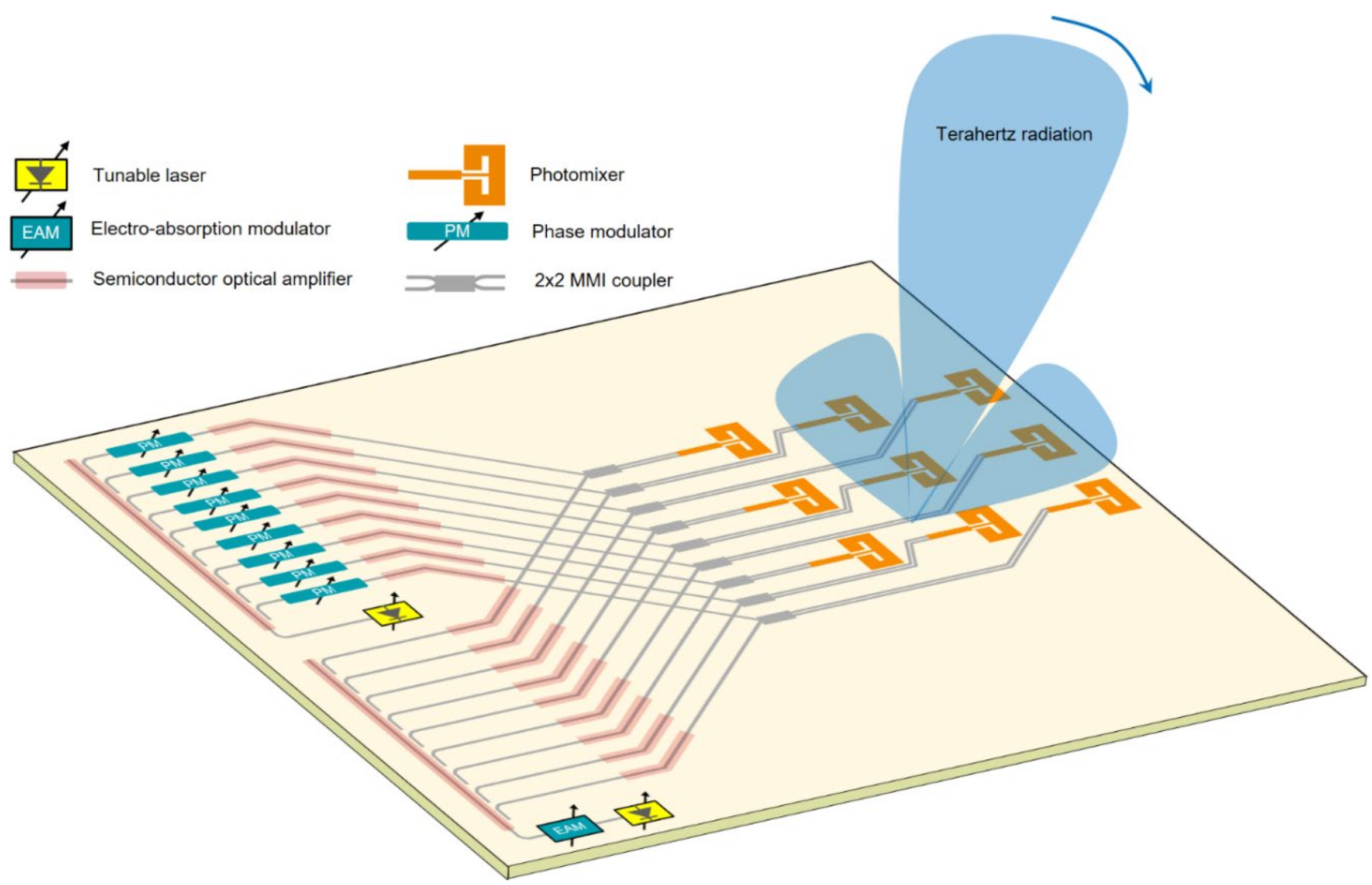

**Fig. 6 | Concept of a single-chip terahertz phased array transceiver based on a QW PIN PIC process for adaptive hyperspectral remote sensing and communication.** A pair of tunable lasers, with an adjustable terahertz frequency difference, is used to generate optical pump beams with a tunable terahertz beat frequency. These optical beams are amplified by one (or multiple) SOAs and distributed across an array of photomixers, which function as both terahertz transmitters and receivers. Depending on the system requirements, phase and intensity modulators are used to control the intensity and phase of the optical beams pumping each photomixer. For example, to implement communication systems, intensity modulators encode the terahertz carrier signal with data to be transmitted. Furthermore, to implement radar, remote sensing, and smart communication systems, which require spatial scanning of the transmitted terahertz radiation, an array of phase modulators dynamically controls the terahertz phase at each photomixer, enabling tilt in the phase front of the transmitted terahertz radiation. The platform also enables the implementation of spectrometry, spectroscopy, and hyperspectral imaging systems, as the radiated terahertz frequency can be easily tuned by adjusting the laser wavelengths. Furthermore, multi-channel terahertz communication systems can be implemented using orthogonal frequency division multiplexing by employing multiple optical pump beams with different terahertz beat frequencies.

Supplementary Materials for

# Terahertz Generation and Detection through Gain-Enhanced Interband Photomixing in Quantum Well Structures

Yifan Zhao[1,2], Shahed-E- Zumrat[1,2], Szu-An Tsao[1,2], Mona Jarrahi[1,2*]

[1]Electrical and Computer Engineering Department, University of California; Los Angeles, 90095, USA.

[2]California NanoSystems Institute, University of California; Los Angeles, 90095, USA.

*Corresponding author. Email: mjarrahi@ucla.edu

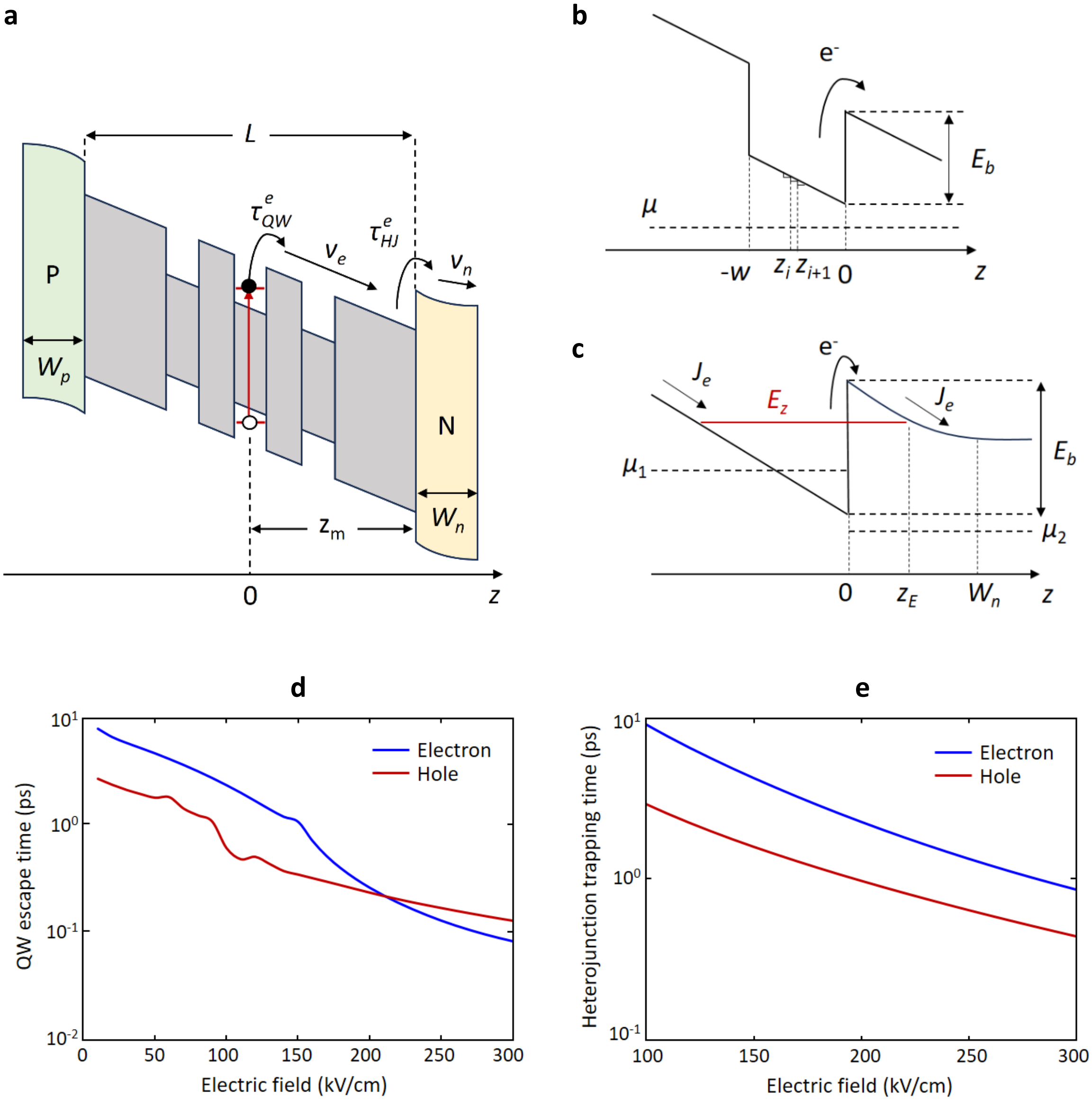

**Fig. S1.** Modeling of ultrafast carrier dynamics in a QW PIN photodiode reproduced from Zhao, Y. and Jarrahi, M., Journal of Applied Physics 139, 015701 (2026). A more detailed analysis is provided in Ref. 29 of the manuscript: Zhao, Y. & Jarrahi, M. Ultrafast carrier dynamics in multiple quantum well p-i-n photodiodes. J. Appl. Phys. 139, 015701 (2026). **a,** Illustration of the PIN structure with multiple QWs embedded in the intrinsic region. Photons are absorbed in the QWs and generate photocarriers. The photo-generated carriers need to escape from the QWs, then drift across the intrinsic region, may be temporarily trapped at the heterojunction interfaces; after escaping the heterojunction barrier, they traverse the depleted cladding layers, and are finally collected at the contact layers. The current generated by the $m^{th}$ QW superimposes with that of other QWs. Thus, the current contributions from all QWs are summed to obtain the total current. We have used the Shockley-Ramo theorem to determine the current induced by the photocarrier motion. The total current is a sum of contributions from electrons and holes in all the QWs. For an optical impulse incident on the structure, the electron density in the $m^{th}$ QW can be expressed as:

$$N(t) = N_m e^{-t/\tau_{QW}^e} u(t) \qquad (S1-1)$$

where $N_m$ is the electron-hole pair density at t = 0 and $\tau_{QW}^e$ is the electron escape time from the QW. After escaping the QW, the carriers drift at a saturation velocity $v_e$, through the intrinsic region with a density:

$$n(z,t) = \frac{N_m}{v_e \tau_{QW}^e} e^{-\left(t-\frac{z}{v_e}\right)/\tau_{QW}^e} u\left(t - \frac{z}{v_e}\right) \qquad (S1-2)$$

where $z$ is the position measured from the center of the QW, $\tau_{trans}^e = \frac{z_m}{v_e}$ is the electron transit time to reach the heterojunction interface, and $z_m$ is the distance between the $m^{th}$ well and the heterojunction interface. According to the Shockley-Ramo theorem, the induced current by electrons moving through the intrinsic region for the $m^{th}$ QW is:

$$I_{e,m,i}(t) = f_i \int_0^{z_m} q v_e S n(z,t)\, dz = \begin{cases} f_i q v_e S N_m \left(1 - e^{-\frac{t}{\tau_{QW}^e}}\right), & 0 < t < \tau_{trans}^e \\ f_i q v_e S N_m \left(1 - e^{-\frac{\tau_{trans}^e}{\tau_{QW}^e}}\right) e^{-\frac{t-\tau_{trans}^e}{\tau_{QW}^e}}, & t \geq \tau_{trans}^e \end{cases} \qquad (S1-3)$$

where $S$ is the device area, $f_i = \frac{1/\varepsilon_i}{L/\varepsilon_i + W_n/\varepsilon_n + W_p/\varepsilon_p}$ is the weighting field in the intrinsic region, $\varepsilon_i$, $\varepsilon_n$ and $\varepsilon_p$ are the dielectric constants of the intrinsic, n-cladding, and p-cladding layers, $W_n$ and $W_p$ are the depletion width of the n- and p-cladding layers, and $L$ is the intrinsic region thickness. The frequency response is obtained by taking the Fourier Transform of this current:

$$I_{e,m,i}(\omega) = \int I_{e,m,i}(t) e^{-j\omega t} dt = z_m f_i \frac{q S N_m}{1 + j\omega \tau_{QW}^e} sinc\left(\frac{\omega \tau_{trans}^e}{2}\right) e^{-\frac{j\omega \tau_{trans}^e}{2}} \qquad (S1-4)$$

This expression represents the induced current for electrons in the $m^{th}$ QW. We can write similar expressions for holes. After the carriers arrive at the heterojunction interface, they can be temporarily trapped due to the barrier height. The rate equation governing the electrons trapped at the heterojunction interface is used to analyze the dynamics of carrier trapping:

$$\frac{dN_t}{dt} = -\frac{N_t}{\tau_{HJ}^e} + n(z_m, t) v_e \qquad (S1-5)$$

where $N_t$ is the trapped electron density, $n(z_m, t)$ is the electron density at the interface as a function of time, and $\tau_{HJ}^e$ is the heterojunction electron trap time. We obtain $N_t(t)$ by solving this equation:

$$N_t(t) = \frac{N_m}{1 - \frac{\tau_{QW}^e}{\tau_{HJ}^e}} \left( e^{-\frac{t-\tau_{trans}^e}{\tau_{HJ}^e}} - e^{-\frac{t-\tau_{trans}^e}{\tau_{QW}^e}} \right) u(t - \tau_{trans}^e) \qquad (S1-6)$$

After the carriers escape the heterojunction barrier, they traverse the depleted cladding region at saturation velocity $v_n$ . The induced current by these electrons moving through the depleted n-cladding is:

$$I_{e,m,n}(t) = f_n q v_n S \int_{t-t_n}^{t} \frac{N_t(t)}{\tau_{HJ}^e} dt \qquad (S1-7)$$

where $t_n = \frac{W_n}{v_n}$ is the electron transit time in the depleted n-cladding layer, $f_n = \frac{1/\varepsilon_n}{L/\varepsilon_i + W_n/\varepsilon_n + W_p/\varepsilon_p}$ is the weighting field of the n-cladding layer. The frequency response is obtained by taking the Fourier Transform of this current:

$$I_{e,m,n}(\omega) = W_n f_n \frac{qSN_m}{1 + j\omega\tau_{QW}^e} \frac{1}{1 + j\omega\tau_{HJ}^e} sinc\left(\frac{\omega t_n}{2}\right) e^{-j\omega\left(\tau_{trans}^e + \frac{t_n}{2}\right)} \qquad (S1-8)$$

Total induced current is the summation of electron and hole currents generated in all QWs:

$$I(\omega) = \sum_m \left[ I_{e,m,i}(\omega) + I_{e,m,n}(\omega) + I_{h,m,i}(\omega) + I_{h,m,p}(\omega) \right] \qquad (S1-9)$$

Here, $I_{h,m,i}$ is the current induced by holes originating from the $m^{\text{th}}$ well moving through the intrinsic region and $I_{h,m,p}$ is the current induced by the holes moving through the depleted p-cladding. **b,** The conduction band profile of a symmetric QW with barrier height $E_b$ and width $w$ under electric field along the $z$-axis. Under an applied electric field $F$, conduction band profile can be written as: $V(z) = -Fz$ within the well ($-w<z<0$) and $V(z) = E_b - Fz$ outside the well ( $z < -w$ or $z > 0$). Under the influence of the electric field, electron states with energy below the barrier become quasi-bounded, allowing for tunneling with a finite probability. The combined tunneling and thermionic emission currents form the total electron escape current, with the escape time constant defined as $\tau_{QW}^e = \frac{N_{QW,e}}{J_e}$, where $J_e$ is the electron escape current density and $N_{QW,e}$ is the 2D electron density in the well. Total escape current is calculated by integrating over all k-space states weighted by the Fermi distribution:

$$J_e = q \iint_{-\infty}^{\infty} \frac{dk_x dk_y}{(2\pi)^2} \left[ \int_{k_{th}}^{\infty} f(E) \langle v_g^+(E_z) \rangle T^+(E_z) \frac{dk_z}{2\pi} \right] \qquad (S1-10)$$

where $f(E) = \frac{1}{e^{(E-\mu)/kT} + 1}$ is the Fermi-Dirac distribution, $\mu$ is the Fermi level, $\langle v_g^+(E_z) \rangle$ is the average group velocity in the $+z$ direction, and $T^+(E_z)$ is the transmission probability at the barrier. By reformulating in terms of energy:

$$J_e = q\int_0^{\infty}\frac{m_e}{2\pi\hbar^2}dE_t\left[\int_{E_{th}}^{\infty}f(E)\langle g_{1d}(E_z)\rangle\langle v_g^+(E_z)\rangle T^+(E_z)dE_z\right] \qquad (S1-11)$$

where $m_e$ is the electron effective mass, $E = E_t + E_z$ is total energy decomposed into in-plane ($E_t$) and perpendicular ($E_z$) components, $\langle g_{1d}(E_z)\rangle$ is the average 1D density of states (DOS), $E_{th} = E_z(k_{th})$, and $k_{th}$ is the threshold below which tunneling is negligible. Below the threshold, the energy states are well confined with discrete, step-like DOS which is similar to an un-biased QW. Above the threshold, states are broadened with continuous DOS and contribute to the escape current. A transmission line analogy is used to compute the 1D DOS, group velocity and transmission probability. The 2D electron density $N_{QW,e}$ is calculated by:

$$N_{QW,e} = w\int_0^{\infty} g_{3d}(E)f(E)dE \qquad (S1-12)$$

Here, $g_{3d}(E)$ is the 3D DOS which varies with energy. For electrons with energy below threshold, discrete DOS is given by:

$$\underset{E<E_{th}}{g_{3d}(E)} = \sum_{E_l<E}\frac{m_e}{\pi^2\hbar^2}(k_l - k_{l-1}) \qquad (S1-13)$$

where $k_l = \sqrt{2m_eE_l}/\hbar$ and $k_0 = 0$. For electrons with energy above the threshold, the continuous DOS is determined by

$$\underset{E>E_{th}}{g_{3d}(E)} = \sum_{E_l<E_{th}}\frac{m_e}{\pi^2\hbar^2}(k_l - k_{l-1}) + \int_0^{\infty}\frac{m_e}{2\pi\hbar^2}dE_t\int_{E_{th}}^{E}\langle g_{1d}(E_z)\rangle\delta(E - E_t - E_z)dE_z \qquad (S1-14)$$

To determine QW escape time for holes, we assumed heavy and light holes are in equilibrium with a common quasi-Fermi level. The escape current is calculated separately for heavy and light holes then summed to obtain the total hole current density. **c,** Conduction band profile at the interface between the intrinsic layer and n-cladding layer. At the heterojunction interface carriers can escape via thermionic emission or tunneling through the barrier. Total current across the interface includes both tunneling and thermionic components. For a heterojunction with a conduction band barrier $E_b$, the conduction band is described by $E_c(z) = -Fz$ for z < 0 and $E_c(z) = E_b + \frac{1}{2}FW_n[1 - \left(1 - \frac{z}{W_n}\right)^2]$ for 0 < z < $W_n$. The electron transmission probability through the barrier is:

$$T(E_z) = \begin{cases} \exp\left(-\frac{2}{\hbar}\int_0^{z_E}[2m_{e2}(E_c(z) - E_z)]^{1/2}dz\right), & if\ E_{min} \le E_z \le E_c(0^+) \\ 1, & if\ E_z \ge E_c(0^+) \end{cases} \qquad (S1-15)$$

where $E_c(z)$ is the conduction band profile, $E_{min}$ = max($E_c(0^-)$, $E_c(W_n)$) is the minimum energy level for electron tunneling, $z$ = 0 is the position of the heterojunction interface, $z_E$ is the position where $E_c(z_E) = E_z$, and $m_{e2}$ is the electron effective mass in the cladding layer. Total current can be determined:

$$J_e = J_{tunn} + J_{therm} = \frac{AT}{k}\int_{E_{min}}^{E_c(0^+)} f'(E_z)T(E_z)\,dE_z + \frac{AT}{k}\int_{E_c(0^+)}^{\infty} f'(E_z)T(E_z)\,dE_z \quad (S1-16)$$

where $A = \frac{4\pi q m_{e1} k^2}{h^3}$ is the Richardson constant, $m_{e1}$ is the effective electron mass on the intrinsic side, and $f'(E_z) = \ln\left(1+\exp[(\mu_1 - E_z)/kT]\right)$ is the Fermi distribution integrated over in-plane states. 2D density of trapped electrons at the heterojunction interface is calculated as:

$$N_{HJ,e} = \int_0^{-\infty} N_c e^{-(E_c(z)-\mu_1)/kT}\,dz \quad (S1-22)$$

where $N_c$ is the effective conduction band density of states in the intrinsic region, $\tau_{HJ}^{e} = \frac{N_{HJ,e}}{J_e}$ is the electron trapping time at the heterojunction, and $\tau_{HJ}^{h} = \frac{N_{HJ,h}}{J_h}$ is the hole trapping time, which is calculated similarly. **d,** The calculated electron and hole escape times from the QWs and **e,** the calculated trapping times at the heterojunction interfaces for the GaAs/AlGaAs QW PIN photodiode heterostructure used for fabricating the terahertz source/detector prototypes. The calculation details are provided in Ref. 29 of the manuscript: Zhao, Y. & Jarrahi, M. Ultrafast carrier dynamics in multiple quantum well p-i-n photodiodes. J. Appl. Phys. 139, 015701 (2026). Under increasing reverse bias, the electric field across the intrinsic region becomes stronger, tilting the potential barriers and enhancing carrier tunneling out of the wells. As a result, the carrier escape time from the QWs ($\tau_{QW}$) becomes shorter at higher bias voltages. Similarly, the carrier transit time ($\tau_{trans}$) decreases with reverse bias because carriers are accelerated more rapidly and quickly reach their saturation velocity, reducing the time required to traverse the depletion region. The heterostructure composition and layer thickness also play a central role. Higher barrier heights set by material composition (e.g., higher Al content in AlGaAs) and thicker barriers increase confinement, leading to longer escape times, while thinner wells and barriers, along with lower band offsets, reduce the escape time by facilitating tunneling. Effective mass differences between materials further influence escape rates by modifying the tunneling probability. The total thickness of the intrinsic region directly determines the transit distance; thus, thicker intrinsic layers increase $t_{trans}$, whereas thinner regions reduce it. In addition, material-dependent saturation velocities and the presence of heterojunction barriers (which can temporarily trap carriers) further modulate the effective transit time. Overall, higher reverse bias and thinner, lower-barrier heterostructures lead to shorter $\tau_{QW}$ and $\tau_{trans}$, while thicker layers and higher band offsets increase these characteristic times.

| Material | Mole Fraction (x) | Thickness (nm) | Type | Doping level ($cm^{-3}$) | Description |
|---|---|---|---|---|---|
| GaAs | | 200 | p-doped(C) | $>2\times10^{19}$ | P-contact |
| Al(x)GaAs | 0.55 to 0.05 | 50 | p-doped(C) | $>3\times10^{18}$ | |
| Al(x)GaAs | 0.55 | 1250 | p-doped(C) | $1\times10^{18}$ | P-cladding |
| GaIn(x)P | 0.49 | 10 | p-doped(Zn) | $7\times10^{17}$ | Etch stop |
| Al(x)GaAs | 0.55 | 300 | p-doped(C) | $6\times10^{17}$ | P-cladding |
| Al(x)GaAs | 0.3 | 40 | Undoped | | Barrier |
| Al(x)GaAs | 0.08 | 5.5 | Undoped | | QW |
| Al(x)GaAs | 0.3 | 6 | Undoped | | Barrier |
| Al(x)GaAs | 0.08 | 5.5 | Undoped | | QW |
| Al(x)GaAs | 0.3 | 6 | Undoped | | Barrier |
| Al(x)GaAs | 0.08 | 5.5 | Undoped | | QW |
| Al(x)GaAs | 0.3 | 40 | Undoped | | Barrier |
| Al(x)GaAs | 0.55 | 1500 | n-doped(Si) | $1\times10^{18}$ | N-cladding |
| Al(x)GaAs | 0.05 to 0.55 | 50 | n-doped(Si) | $2\times10^{18}$ | N-cladding |
| GaAs | | 1000 | n-doped(Si) | $2\times10^{18}$ | N-contact |
| SI-GaAs | | | Undoped | | Substrate |

**Fig. S2.** The GaAs/AlGaAs QW PIN photodiode heterostructure used for fabricating the terahertz source/detector prototypes is a commercially available design typically utilized in semiconductor lasers operating at an approximate wavelength of 800 nm. This structure was grown by Xiamen Powerway Advanced Material Co., Ltd (PAM-XIAMEN). From bottom to top, the structure comprises: a 1-µm-thick, highly doped n+ GaAs contact layer grown on a semi-insulating (SI) GaAs substrate; a 50-nm-thick graded AlGaAs layer, followed by a 1.5-µm-thick AlGaAs n-cladding layer; a 108.5-nm-thick intrinsic region; a 1.5-µm-thick AlGaAs p-cladding layer, which includes a 10-nm-thick lattice-matched GaInP etch stop layer positioned 300 nm above the intrinsic region; and another 50-nm-thick graded AlGaAs layer and a 200-nm-thick, highly doped p+ GaAs contact layer. Within the intrinsic region, there are three pairs of QWs, each consisting of a 5.5-nm-thick $Al_{0.08}Ga_{0.92}As$ well layer and a 6-nm-thick $Al_{0.3}Ga_{0.7}As$ barrier layer, flanked by a 40-nm-thick $Al_{0.3}Ga_{0.7}As$ layer on each side. Although this wafer structure enables high-performance terahertz transmitter and receiver functionality, there remains potential for optimization to enhance its suitability for integrated terahertz optoelectronic applications.

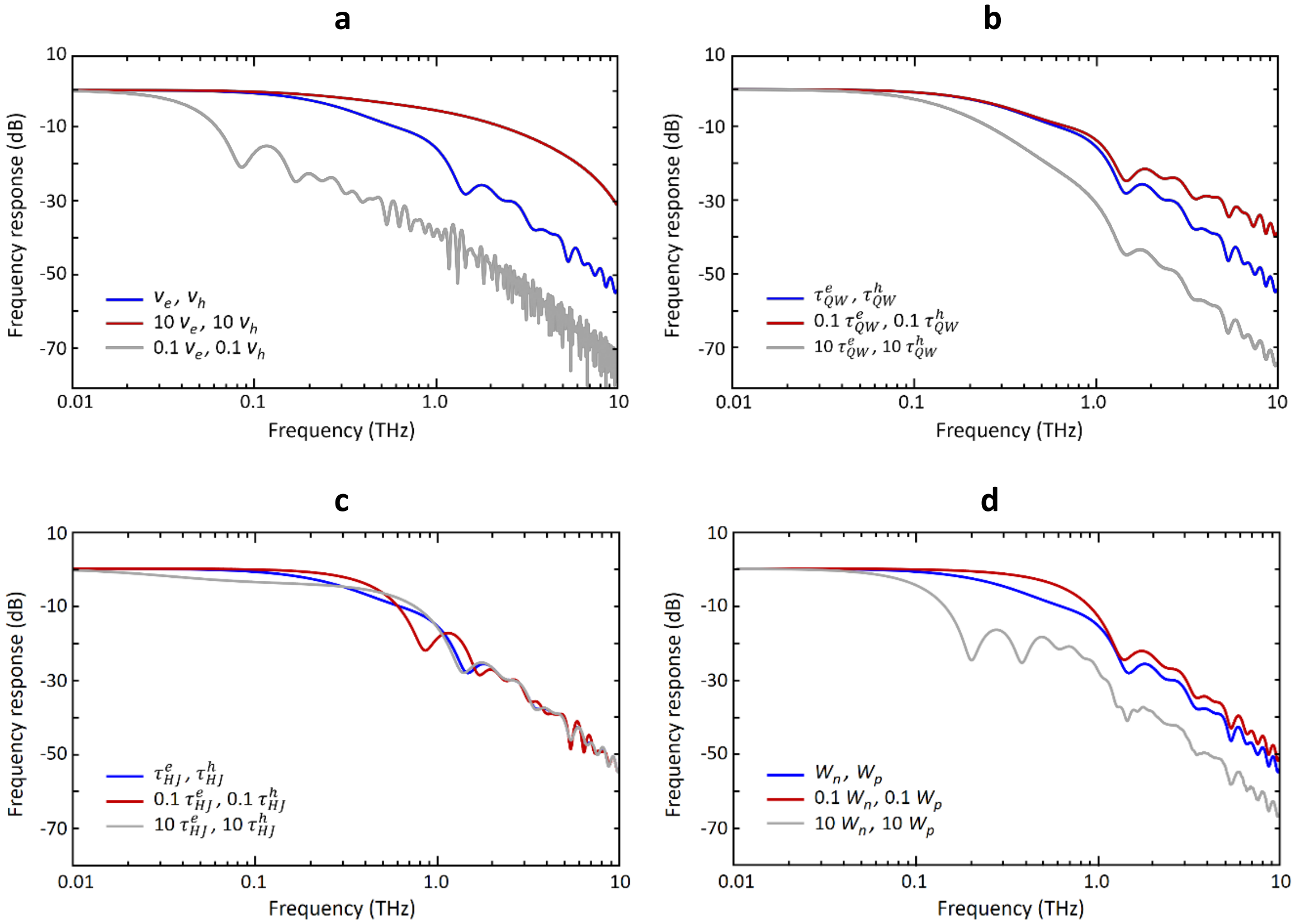

**Fig. S3.** Frequency response of the GaAs/AlGaAs QW PIN photodiode heterostructure used for fabricating the terahertz source/detector prototypes reproduced from Zhao, Y. and Jarrahi, M., Journal of Applied Physics 139, 015701 (2026). This frequency response is calculated using the model described in Fig. S1 and includes the impact of the electron and hole escape times from the QWs ($\tau_{QW}^e$ and $\tau_{QW}^h$), electron and hole transit times through the intrinsic region ($\tau_{trans}^e$ and $\tau_{trans}^h$), electron and hole trap times at the heterojunction interface ($\tau_{HJ}^e$ and $\tau_{HJ}^h$), and excludes the impact of the device RC time constant, which is described in Figs. S4 and S5. The blue plot illustrates the frequency response based on the electron/hole saturation velocities of $0.72\times10^7$ / $0.8\times10^7$ cm/s through the depletion region [30, 31], as well as the calculated QW electron/hole escape times of 0.09/0.13 ps and heterojunction electron/hole trap times of 0.98/0.48 ps at an applied field of 280 kV/cm (as shown in Fig. S1 d, e). For frequencies between 0.1 and 1 THz, the frequency response exhibits a 20 dB/decade slope due to the slow transit time of ~1 ps. At higher frequencies, the slope increases to 40 dB/decade as the QW escape time (~0.1 ps) introduces an additional bottleneck. The calculation details are provided in Ref. 29 of the manuscript: Zhao, Y. & Jarrahi, M. Ultrafast carrier dynamics in multiple quantum well p-i-n photodiodes. J. Appl. Phys. 139, 015701 (2026). **a,** The impact of carrier transit time on the frequency response is studied by varying the carrier saturation velocity, while maintaining the geometry of the QW PIN structure,

illustrating the $sinc^2\left(\frac{\omega t_{trans}}{2}\right)$ dependence predicted by the theoretical model described in Fig. S1. Reducing the thickness of the intrinsic region while maintaining the same QW structures would decrease the carrier transit time and reduce the frequency roll-off. **b,** The impact of QW escape time on the frequency response is studied, illustrating the $(1 + j\omega\tau_{QW})^{-1}$ dependence predicted by the theoretical model. **c**, The impact of heterojunction trap time on the frequency response is studied, indicating a negligible dependence. This is because the heterojunction trap time only affects the current contribution from the photocarriers moving inside the depleted n- and p-cladding layers with depletion widths $W_n$ and $W_p$, respectively. In the structure studied, the doping level is high enough to keep the depletion width in the cladding layer relatively small compared to the intrinsic region, resulting in minimal effect on the frequency response. **d,** The impact of cladding layer doping on the frequency response is studied. Using lighter doping in the cladding layers extends the depletion width, which increases the carrier transit time and results in a steeper frequency roll-off.

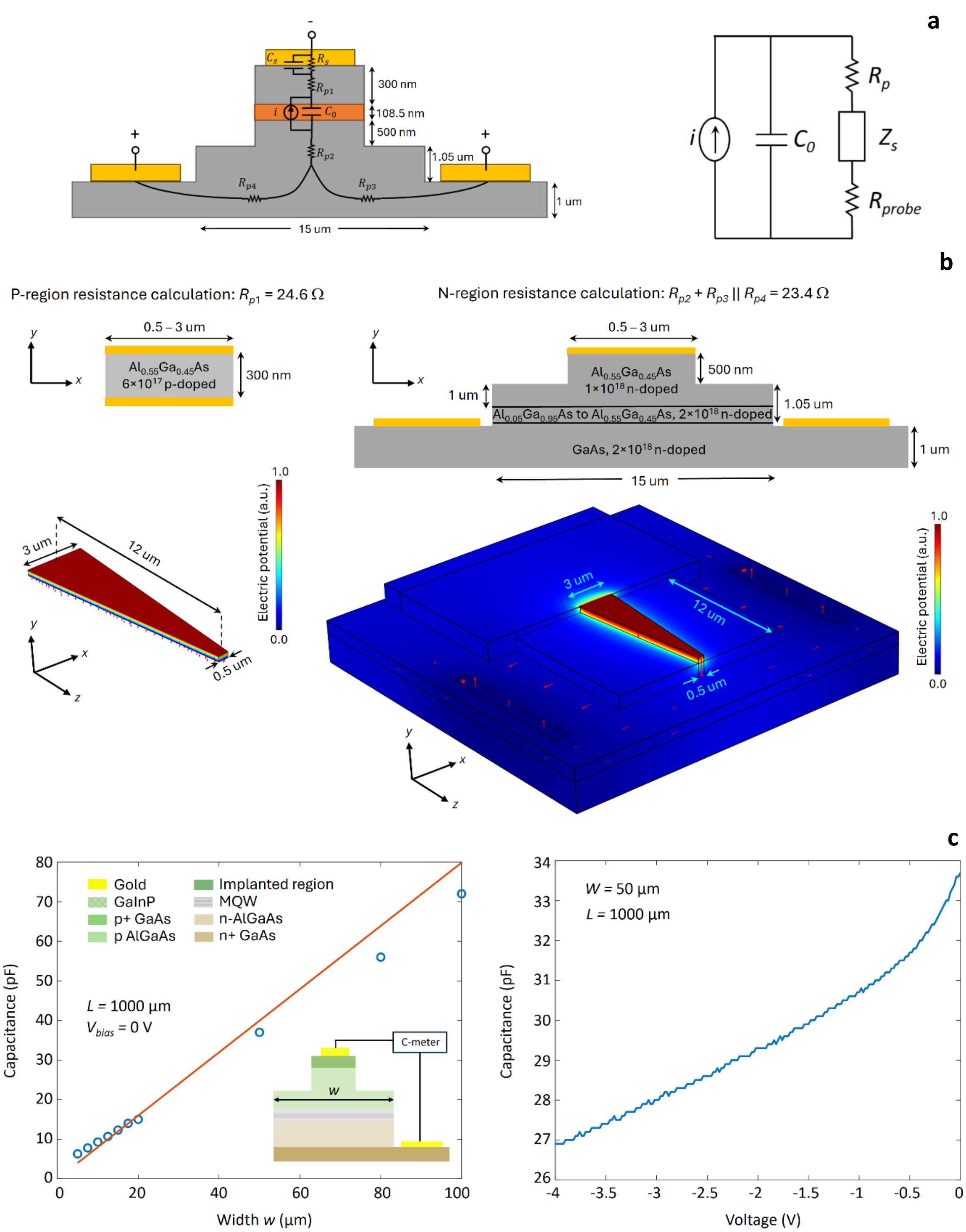

**Fig. S4. a,** The equivalent circuit model for the photomixer in terahertz generation mode is represented with the generated photocurrent modeled as a current source $i = I_{DC} + i_{THz}\cos(2\pi f_{beat}t)$, where $I_{DC}$ and $i_{THz}$ are the amplitudes of the DC and terahertz components of the photocurrent. Here, $f_{beat}$ represents the terahertz beat frequency of the two optical tones

pumping the photomixer. The terahertz photocurrent component is given by $i_{THz} \cong i_{DC}.\, sinc(\pi f_{beat}\tau_{trans})\left(1 + j2\pi f_{beat}\tau_{QW}\right)^{-1}$, where $\tau_{trans}$ is the effective carrier transit time from the QWs to the P/N layers, and $\tau_{QW}$ is the effective carrier escape time from the QWs. The current source is in parallel with the PIN diode depletion region capacitance $C_0$ and this combination is in series with the 50 Ω resistance of the GSG probe, $R_{probe}$, the parasitic resistance $R_p$ from the n and p regions, and the Schottky contact impedance $Z_s$. The terahertz current passing through the load resistance is thus: $i_L = i_{THz}\left[1 + j2\pi f_{beat}C_0(R_{probe} + R_p + Z_s)\right]^{-1} \equiv i_{THz}[1 + j2\pi f_{beat}\tau_{RC}]^{-1}$ and the generated terahertz power is $\frac{1}{2}\left|i_L^2\right|R_{probe}$, where $\tau_{RC}$ represents the photomixer RC time constant for the designed photomixer shown in Fig. 2. **b,** The parasitic resistance $R_p$ from the n and p regions is calculated using COMSOL. The DC current flowing through the p and n regions is simulated separately under a DC voltage and the corresponding resistance values for the p and n regions are calculated as $R_{p1}$ = 24.6 Ω and $R_{p2} + R_{p3} \,||\, R_{p4}$ = 23.4 Ω, with a total parasitic resistance of $R_p$ = 48 Ω. **c**, The PIN diode depletion region capacitance $C_0$ is determined experimentally. Due to the small photomixer active area, it is not feasible to directly measure the capacitance with a capacitance meter. Instead, we fabricated 1-mm-long ridge waveguide samples with the cross-section shown in the inset and measured their capacitance using an HP 4280A capacitance meter. The width, *w*, in these samples varies from 5 to 100 μm across different samples. The left graph shows the measured capacitance values (blue data points) as a function of width at a 0 V bias voltage. The results are in close agreement with the estimated parallel plate capacitance (red curve), $\varepsilon Lw/t$, where $\varepsilon \approx 12\varepsilon_0$ is the average permittivity of the PIN depletion region, *L* = 1 mm, and t is the calculated depletion region thickness at 0 V bias (133 nm). The right graph shows the measured capacitance values as a function of reverse bias voltage for a device with a width of 50 μm. The device capacitance decreases with increasing the reverse bias voltage as the depletion region thickness increases. Finally, we extract the capacitance value of $C_0$ = 11.8 fF for our photomixer at a 3 V reverse bias by scaling the experimentally measured capacitance values according to the photomixer active area.

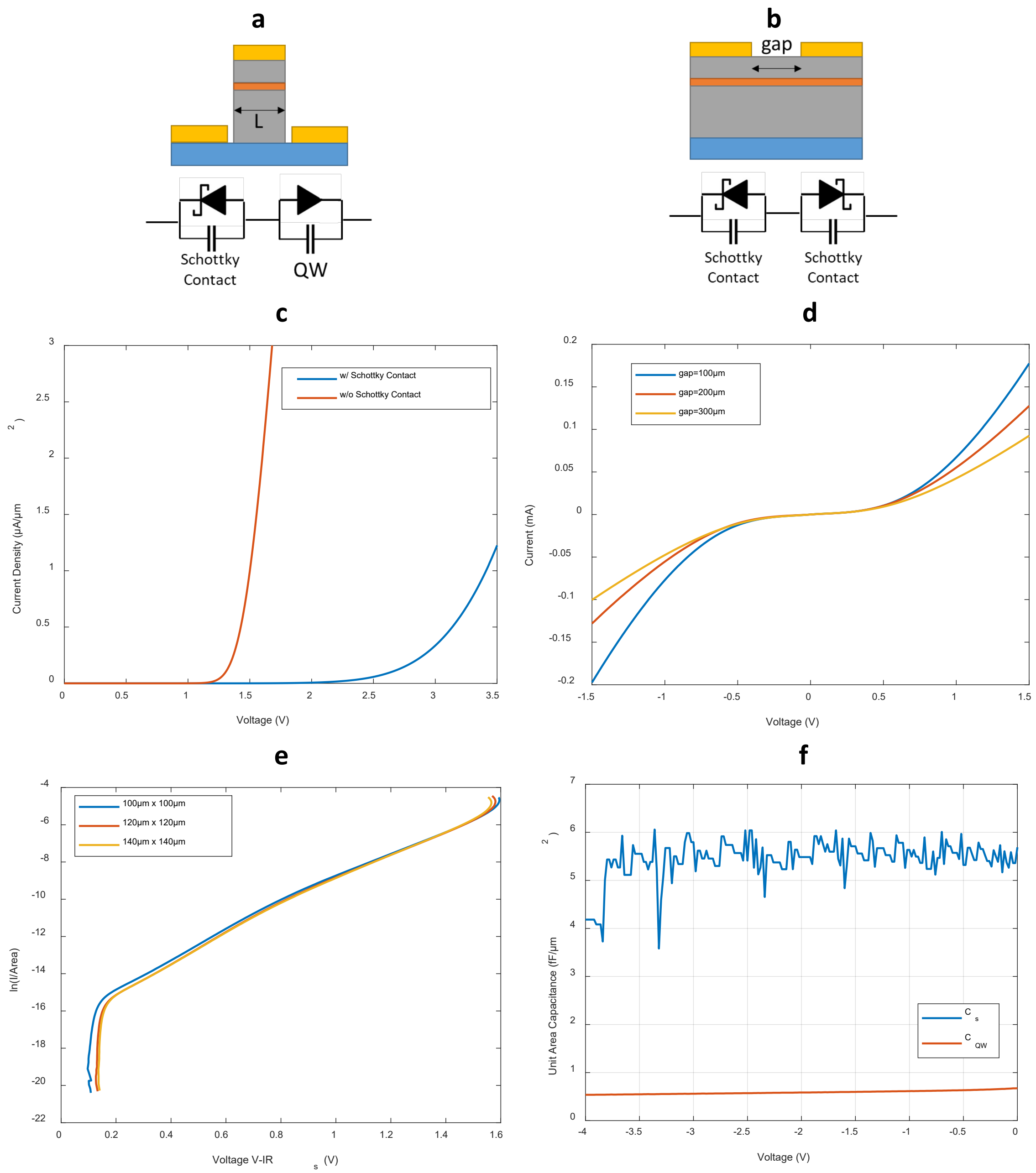

**Fig. S5.** Schottky contact circuit model: Since we etched away the p-type layers above the etch stop layer and deposited the photomixer top contact directly on the etch stop layer to minimize the parasitic resistance of the photomixer, the Schottky contact impedance plays a role in the frequency response of the photomixer. We extracted the capacitance and resistance of the formed Schottky contact using specific test structures shown in **a** and **b**. For the first test structure, we deposited 10/300 nm Cr/Au on the etch stop layer and etched mesas of various sizes. For the bottom contact, we used the same AuGe/Au ohmic contact as in the photomixer. The equivalent circuit model for this test structure is the series connection of the QW diode and the Schottky diode. The second test

structure consists of two Schottky contacts with different gap sizes on a mesa with a 200 μm width. Since the n-side is isolated from the p-side by the depletion region, the equivalent circuit model includes two Schottky diodes connected back-to-back. Both test structures were annealed at 380°C for 30 seconds, a requirement for the AuGe/Au ohmic contact. The IV characteristics of the two test structures are shown in **c** and **d**. When the two Schottky diodes are connected back-to-back, the threshold voltage of the entire structure is dominated by the reverse-biased diode. A threshold voltage of approximately 0.7 V is calculated from the IV characteristics of the second test structure. As illustrated in the IV characteristics of the first test structure, adding the Schottky contact to the QW PIN diode increases the turn-on voltage of the entire structure by approximately 1 V and reduces the current density. This behavior is expected due to the opposite polarity of the Schottky contact relative to the QW diode. As a result, the Schottky contact will be forward-biased when the photomixer is reverse-biased, which has a minor impact on the photomixer operation. To extract the IV characteristics of the Schottky contact, we used the slope of the IV curve to calculate the parasitic resistance and subtracted the IR voltage drop for the first test structure and a pure QW diode with a top ohmic contact. This approach allows us to isolate the IV characteristics of a pure QW diode with a top ohmic contact and a QW diode with a top Schottky contact. By subtracting the voltage across the QW diode at the same current density, we calculated the voltage across the Schottky contact and extracted the IV characteristics of the Schottky contact. The extracted IV characteristics of the Schottky contact from test structures with three different mesa sizes are shown in **e**, demonstrating excellent agreement across all results. To account for the barrier lowering effect under large reverse bias voltages, a Schottky contact can be modeled as:

$$I = I_0 e^{\frac{eV}{nkT}} \left[1 - e^{-\frac{eV}{kT}}\right] \qquad (S5-1)$$

where $I_0$ is the reverse-bias saturation current, $V$ is the applied voltage, $k$ is the Boltzmann constant, $T$ is the temperature, and $n$ is the ideality factor. When $|V| \gg kT/e$, we have

$$I = I_0 \exp\left(\frac{\left(\frac{1}{n} - 1\right) eV}{kT}\right) \qquad (S5-2)$$

$$ln\, I = ln\, I_0 + \frac{e(V - IR_s)}{kT}\left(\frac{1}{n} - 1\right)$$

Therefore, $I_0$ and $n$ can be extracted from the results above and we calculate $I = \frac{5.7pA}{\mu m^2}\left[\exp\left(\frac{eV}{1.27kT}\right) - 1\right]$ for the Schottky contact. Under a DC photocurrent of 1 mA, this corresponds to an AC resistance of $R_s = dV/dI \approx 32.8\Omega$, which is independent of the device area. To compare this with the case where the Schottky contact is absent — meaning the p-type layers above the etch stop are not removed, and the p+ GaAs layer still has an ohmic contact — we calculated the ohmic resistance of the p-type layers to be 60 Ω, which is approximately twice the resistance of the Schottky contact. We also characterized the Schottky contact capacitance by measuring the capacitance of the first test structure under different reverse bias voltages. The measured capacitance corresponds to the series combination of the Schottky contact capacitance, $C_s$, and the depletion region capacitance, $C_{QW}$, given by the equation $C = 1/(\frac{1}{C_{QW}} + \frac{1}{C_s})$. Since the Schottky junction is forward-biased when the QW diode is reverse-biased, most of the reverse bias voltage drops across the depletion region capacitance. Using the depletion region capacitance determined in Fig. S4, we solved for the Schottky contact capacitance $C_s$. The results, shown in **f**, reveal a constant Schottky contact capacitance of 5.5 fF/μm² as a function of the applied voltage, as theoretically expected. To summarize, the Schottky contact impedance is given by:

$$Z_s = \frac{32.8\Omega}{1 + j2\pi f_{beat} \times 32.8\Omega \times 5.5\text{fF} \cdot \mu\text{m}^{-2} * Area(\mu\text{m}^2)} \quad (S5-3)$$

Using the extracted Schottky contact characteristics and the capacitance/resistance parasitic values $C_0 = 11.8$ fF and $R_p = 48$ Ω calculated in Fig. S4, the effective RC time constant of the photomixer is calculated as $\tau_{RC} \approx 1.55$ ps, where $1 + j2\pi f_{beat}\tau_{RC} \approx 1 + j2\pi f_{beat} C_0(R_{probe} + R_p + Z_s)$ in the 0.1-0.5 THz frequency range used in this work.

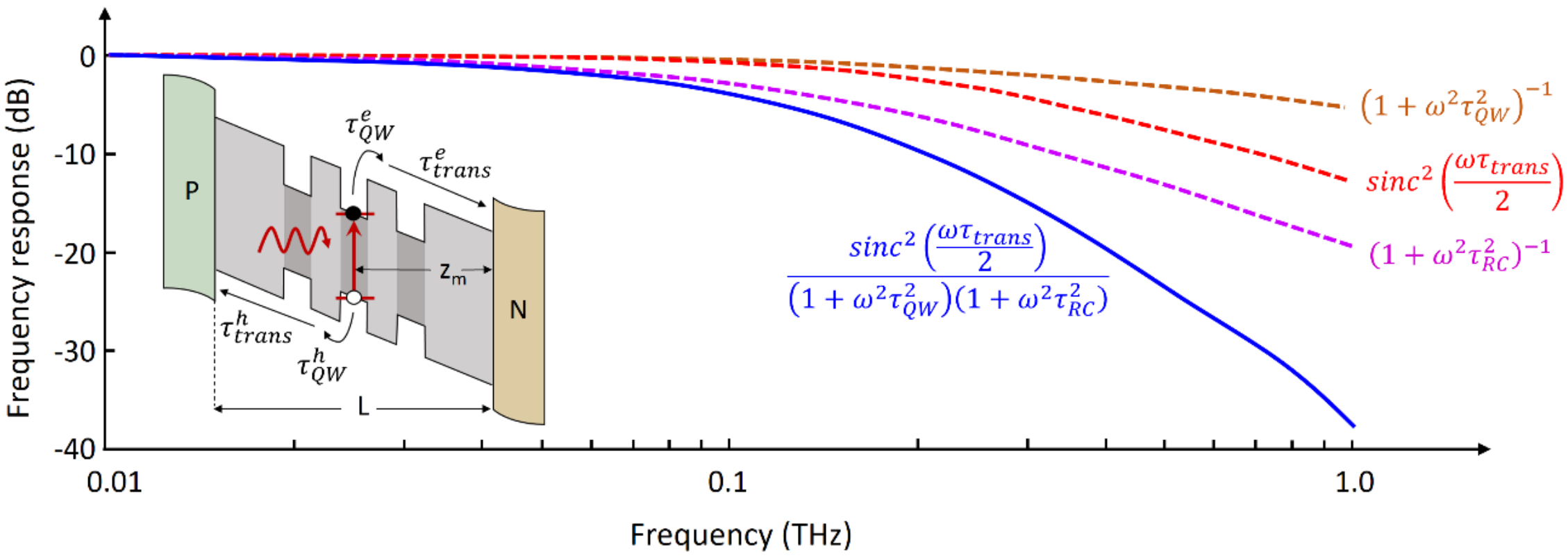

**Fig. S6.** The theoretically predicted frequency response of the QW PIN photomixer, which accounts for the ultrafast carrier dynamics (calculated in Figs. S1 and S3) and the photomixer RC time constant (calculated in Figs. S4 and S5).

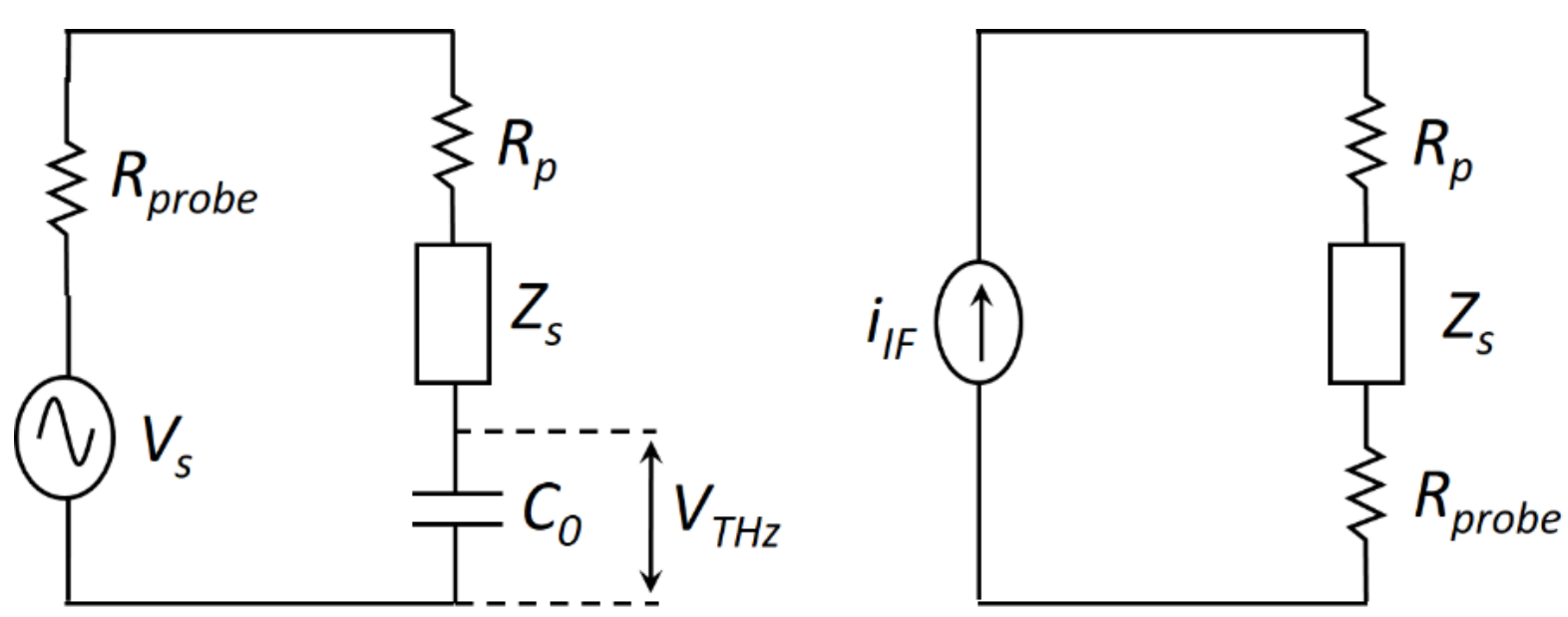

**Fig. S7.** The equivalent circuit model for the photomixer in terahertz detection mode includes a voltage source $V_s$ and the 50 Ω resistance of the GSG probe, $R_{probe}$, to represent the terahertz signal input. These components are in series with the parasitic resistance $R_p$ from the n and p regions, the Schottky contact impedance $Z_s$, and the capacitance $C_0$ of the PIN diode depletion region. For a received terahertz power $P_{THz} = \frac{V_s^2}{8R_{probe}}$, the induced voltage across the intrinsic region is calculated as $V_{THz} = V_s\left[1 + j2\pi f_{beat}C_0(R_{probe} + R_p + Z_s)\right]^{-1}$. As described previously, the photocurrent of the photomixer has both DC and terahertz components, represented by $i = I_{DC} + i_{THz}\cos(2\pi f_{beat}t)$, where $i_{THz} \cong i_{DC}.\,sinc(\pi f_{beat}\tau_{trans})\left(1 + j2\pi f_{beat}\tau_{QW}\right)^{-1} = i_{DC}.\,H_{carrier}(2\pi f_{beat})$. In terahertz detection mode, the voltage applied to the intrinsic region, $V_{DC} + V_{THz}\cos(2\pi f_{THz}t)$, consists of a DC component from the device's DC bias and a terahertz component, $V_{THz}$, induced across the intrinsic region of the photomixer at frequency $f_{THz}$. Assuming $V_{THz} \ll V_{DC}$, the induced photocurrent is given by $i = I_{DC} + I_{DC} * H_{carrier}(V_{DC} + V_{THz}\cos(2\pi f_{beat}t)) * \cos(2\pi f_{THz}t)$, approximated as $I_{DC} + I_{DC} * [H_{carrier}(V_{DC}) + H'_{carrier}(V_{DC}) * V_{THz}\cos(2\pi f_{THz}t)] * \cos(2\pi f_{beat}t)$, yielding an intermediate frequency (IF) component of $\frac{1}{2}I_{DC} * H'_{carrier}(V_{DC}) * V_{THz}\cos|2\pi f_{beat}t - 2\pi f_{THz}t|$. Thus, the magnitude of the IF current is $i_{IF} = \frac{1}{2}I_{DC}V_s * H'_{carrier}(V_{DC})\left[1 + j2\pi f_{beat}C_0(R_{probe} + R_p + Z_s)\right]^{-1}$. The conversion gain of the terahertz detector, defined as the down-converted IF power, $\frac{1}{2}i_{IF}^2R_{probe}$, divided by the received terahertz power, $P_{THz}$, is given by $[I_{DC}R_{probe} * H'_{carrier}(V_{DC})]^2\left[1 + j2\pi f_{beat}C_0(R_{probe} + R_p + Z_s)\right]^{-2}$, where $H'_{carrier}(V_{DC})$is proportional to $sinc(\pi f_{beat}\tau_{trans})\left(1 + j2\pi f_{beat}\tau_{QW}\right)^{-1}$.

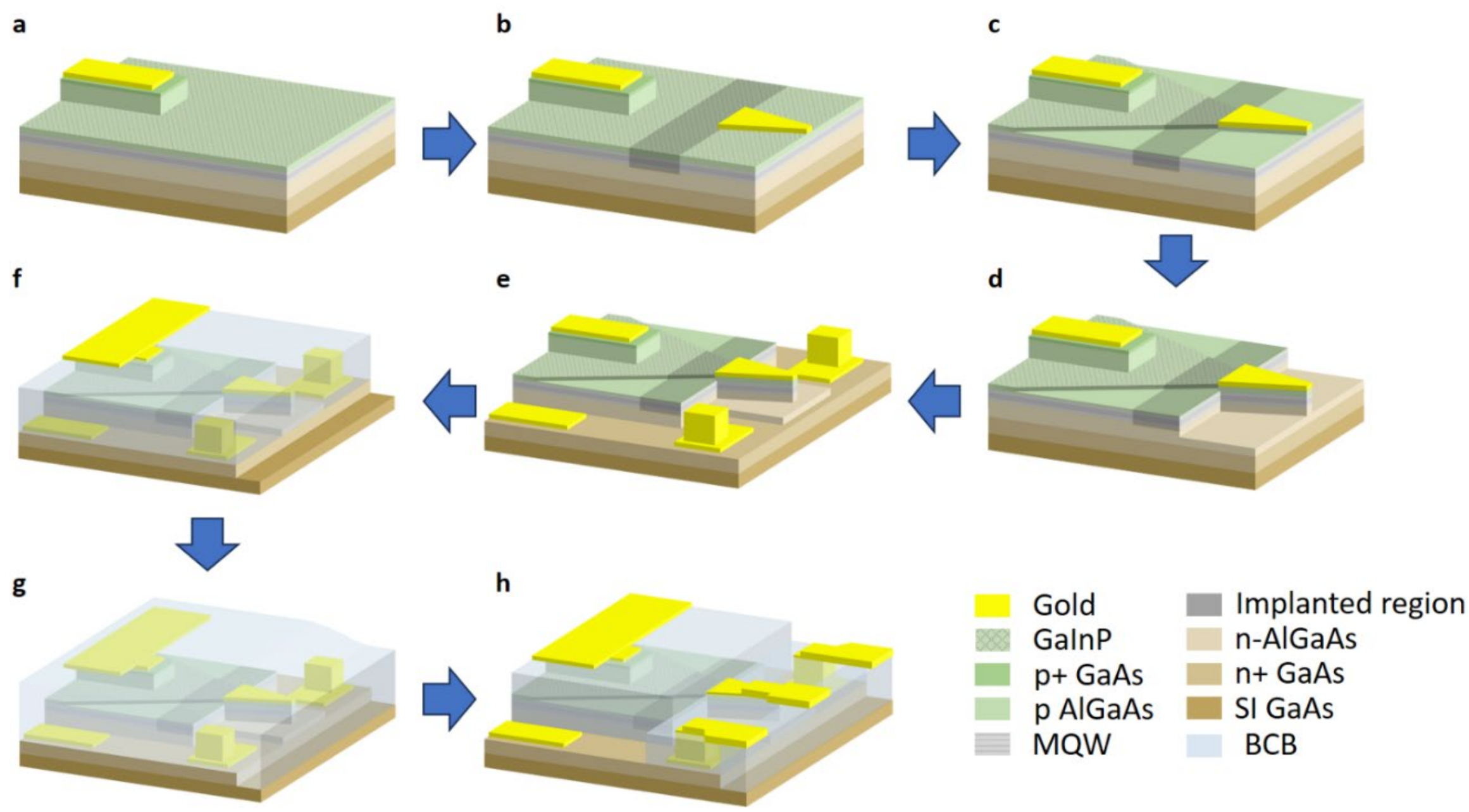

**Fig. S8.** The fabrication process for the terahertz source/detector prototype, which generates and detects terahertz signals via photomixing in a GaAs/AlGaAs QW PIN photodiode, involves the following steps: **a**, Deposition of the SOA top contact and etching of the SOA waveguide. **b**, Deposition of the photomixer top contact followed by ion implantation. **c**, Etching of the shallow taper. **d**, Etching of the photomixer waveguide. **e**, Exposure of the n+ GaAs layer and deposition of the bottom contacts. **f**, Spin coating with BCB (benzocyclobutene), followed by etch-back of BCB to allow deposition of the connection pad for the SOA top contact, and etching of the n+ GaAs in the GSG (ground-signal-ground) pad area. **g**, Spin coating with BCB again. **h**, Etch-back of BCB, deposition of GSG probing pads, and exposure of the SOA contact.

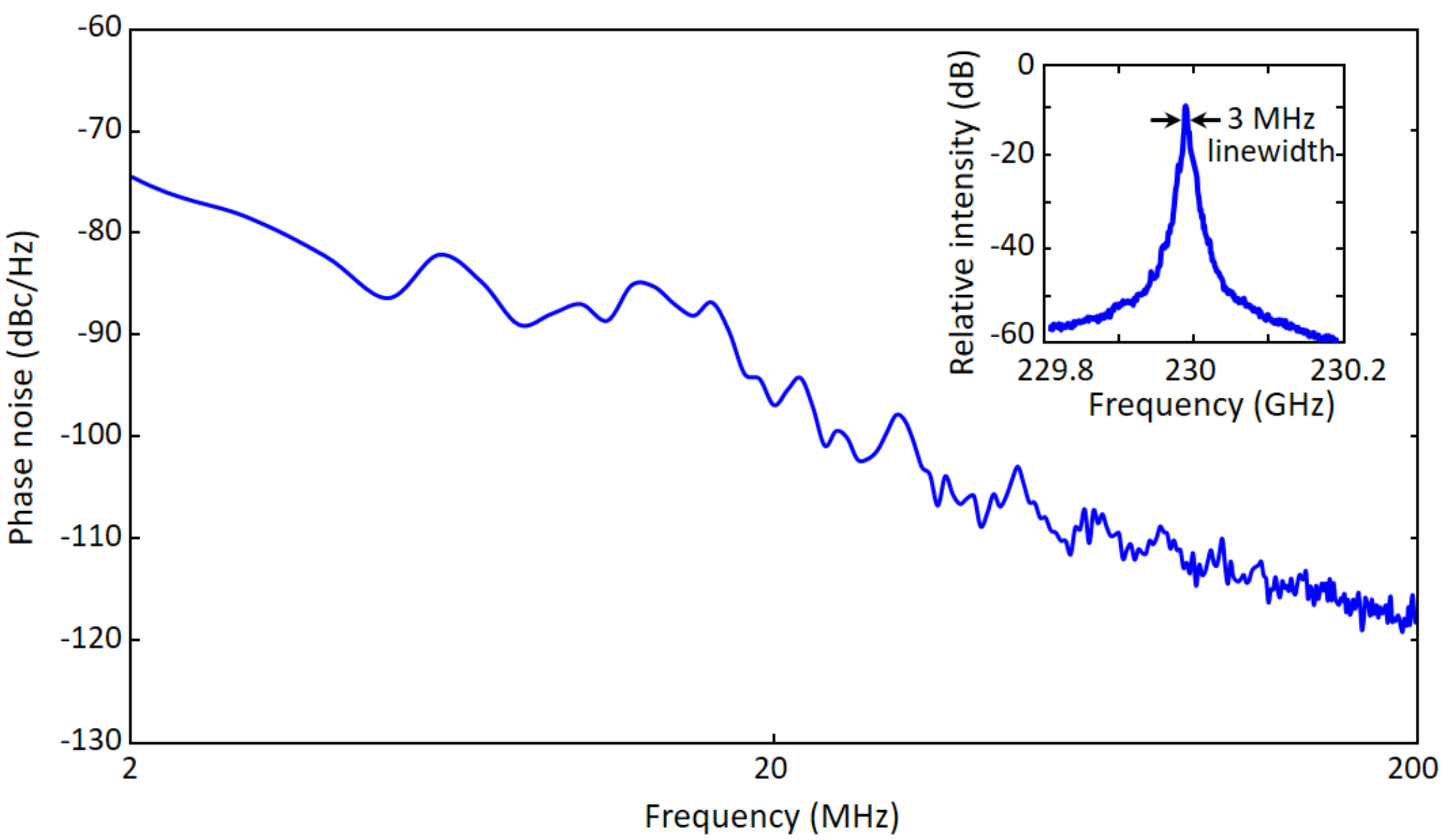

**Fig. S9.** Phase noise of the generated signal at 230 GHz, calculated from the measured signal spectrum shown in the inset.

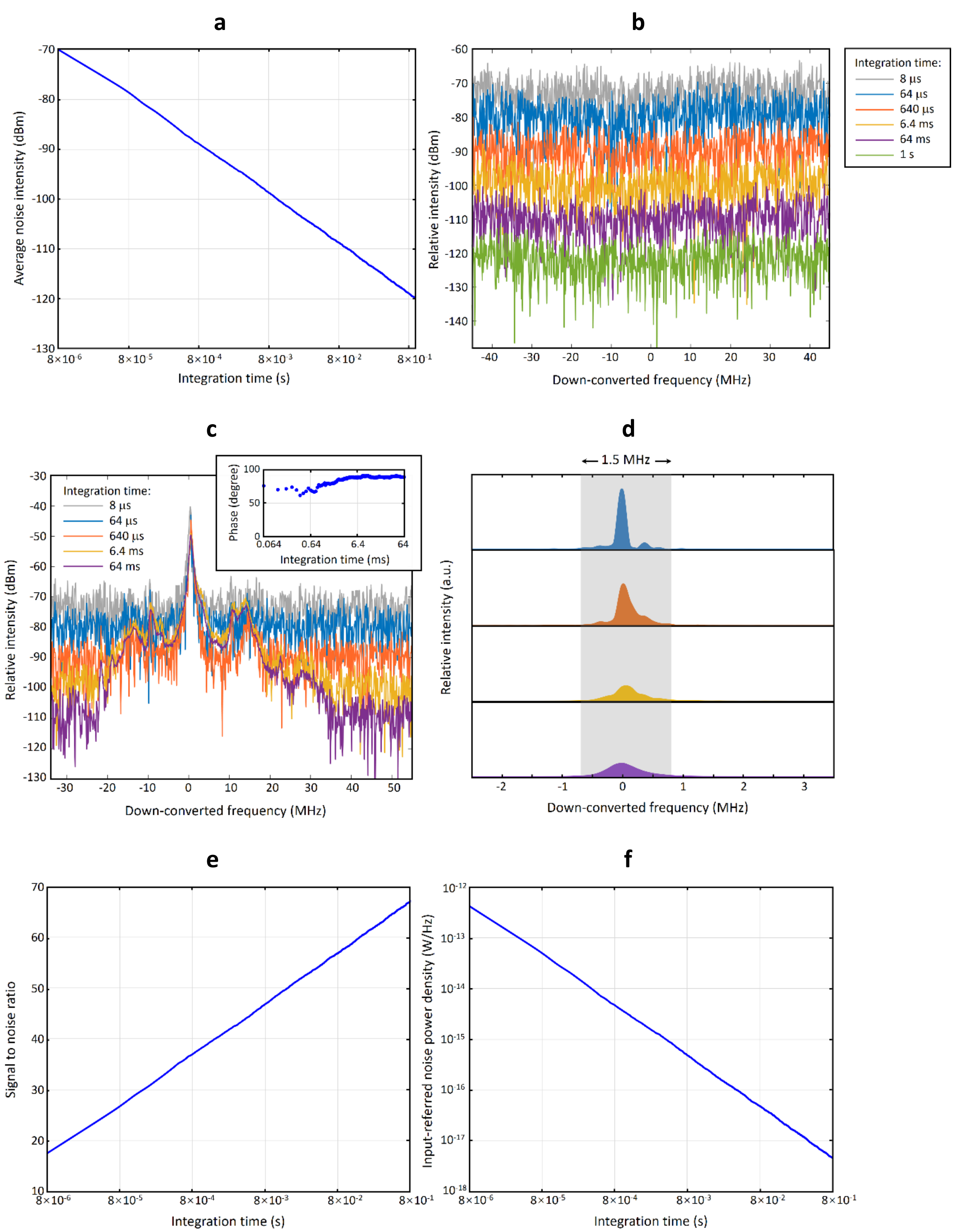

**Fig. S10.** Characteristics of the down-converted terahertz spectrum using spectral lock-in detection. **a,** Average noise intensity of the down-converted spectrum as a function of integration time, showing a noise reduction rate of 10 dB per decade. **b,** Down-converted

spectrum in the absence of a terahertz input for integration times ranging from 8 μs to 1 s. The 8 μs data correspond to measurements captured over an 8 μs time window without applying lock-in detection. **c,** Down-converted spectrum of a 240 GHz input tone at –14.5 dBm for integration times from 8 μs to 64 ms. Inset shows the phase of the resolved spectra at 240 GHz through spectral lock-in detection as a function of integration time. **d,** Down-converted signal power within a 1.5 MHz detection bandwidth remains constant for integration times up to 64 ms, despite IF spectral broadening caused by pump laser wavelength fluctuations. **e,** Signal-to-noise ratio (SNR) of the down-converted 240 GHz tone at –14.5 dBm, calculated as the ratio of signal power to noise power within a 1.5 MHz bandwidth, for integration times ranging from 64 μs to 1 s. **f,** Calculated input-referred noise power density, defined as the input power required to yield an SNR of 1 in a 1 Hz bandwidth ($\frac{-14.5\ dBm - SNR}{1.5\ MHz}$).

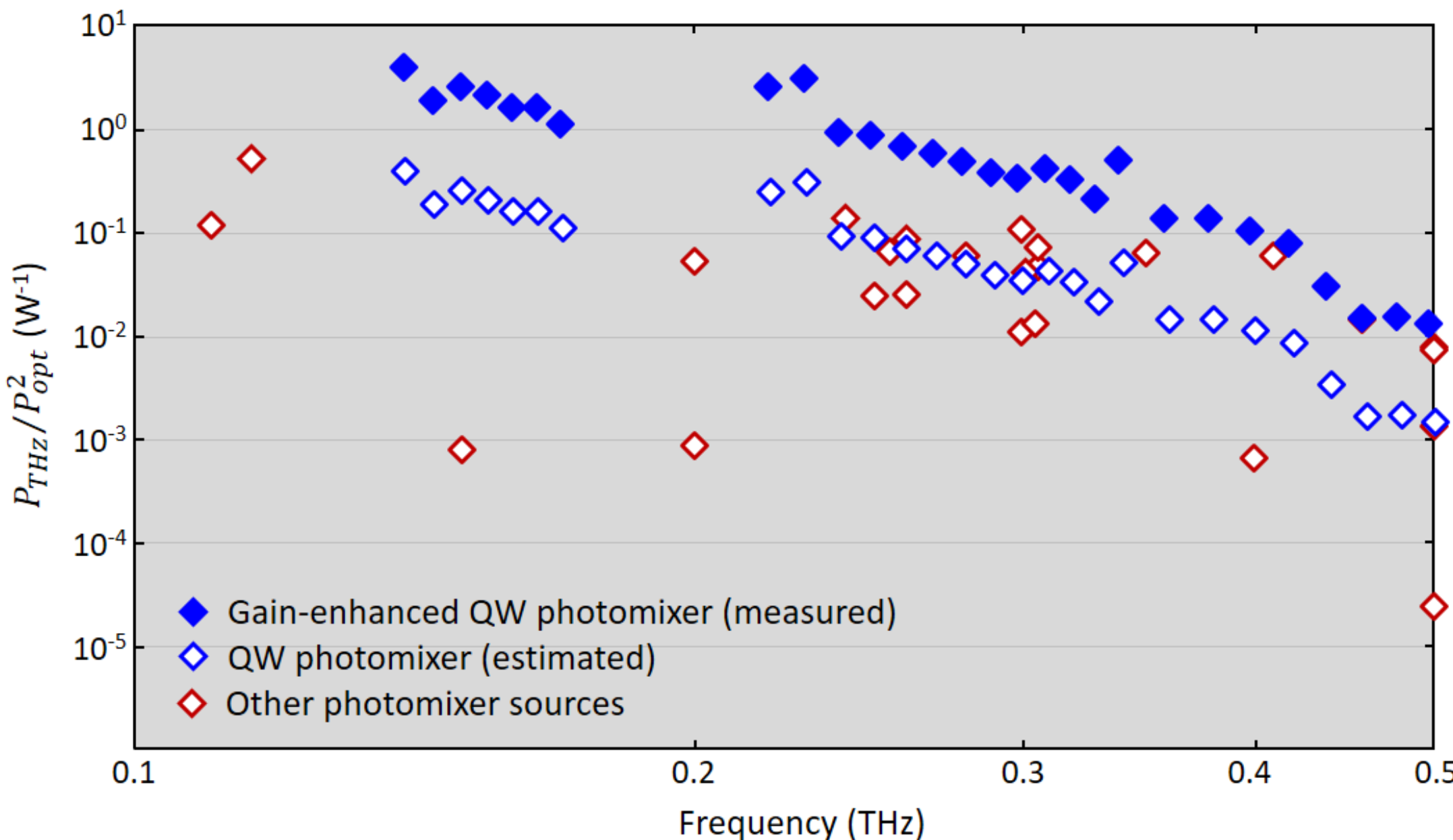

**Fig. S11.** The impact of the SOA gain on the terahertz generation efficiency is evaluated by comparing the efficiency figure of merit $P_{THz}/P^2_{opt}$ at the measured input optical power to the SOA (5 mW) and the estimated input optical power to the photomixer. The estimated input optical power to the photomixer is calculated as $\frac{1}{\eta}\frac{hc}{q\lambda}I_{ph} = 16.3$ mW, where $h$ is the Planck's constant, $c$ is the speed of light, $q$ is the electron charge, $\lambda$ is the optical wavelength of 809 nm, $\eta$ is the photomixer quantum efficiency, estimated to be 56.5% from the electromagnetic simulations (Fig. 2b), and $I_{ph}$ = 6 mA is the photomixer photocurrent. Based on these estimates, the SOA gain is found to enhance the terahertz generation efficiency by approximately an order of magnitude.

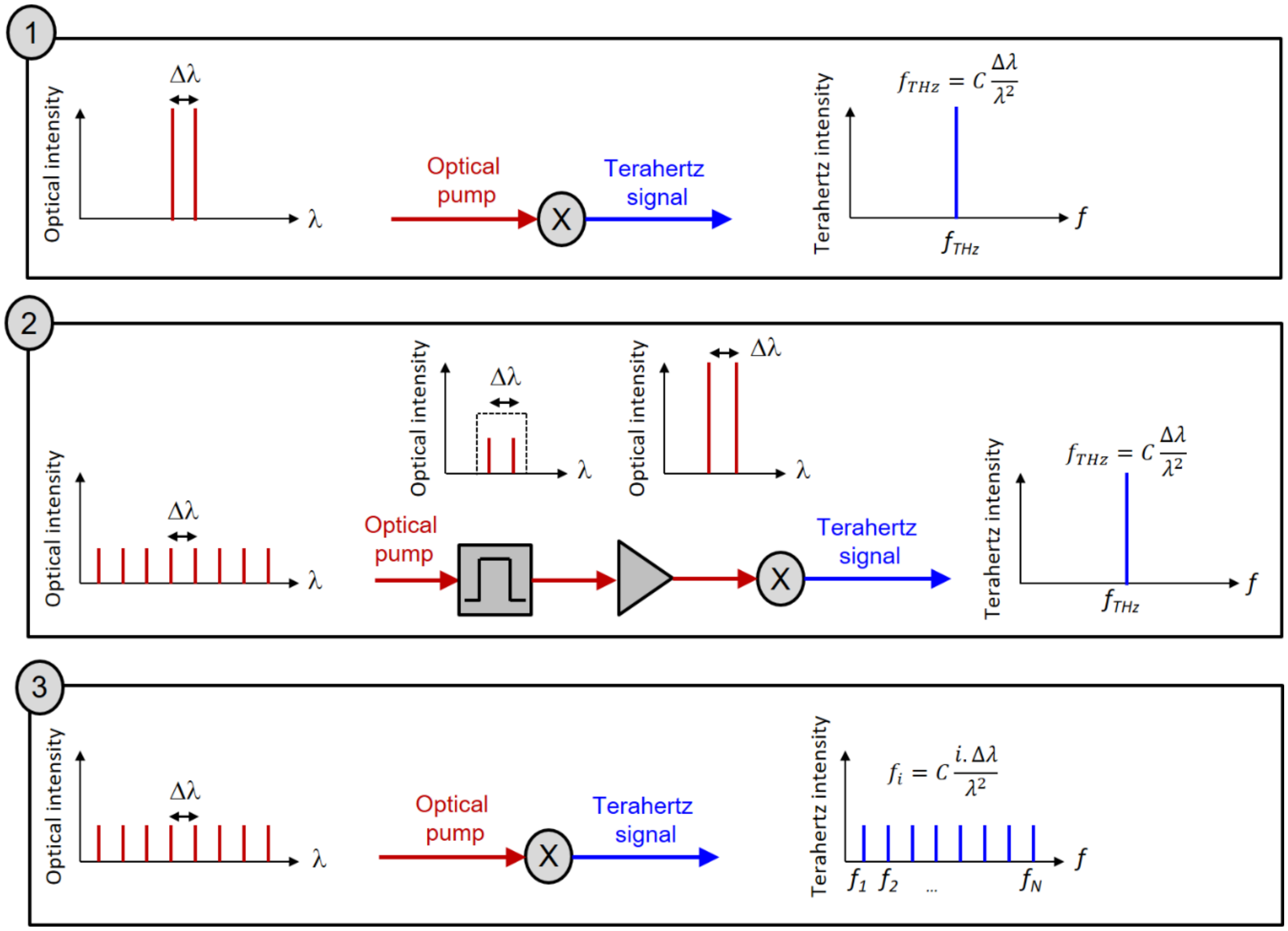

**Fig. S12.** Possible configurations for terahertz generation via photomixing with two-tone and multi-tone optical pump beams. (1) Two-Tone Optical Pumping: When two optical tones with a terahertz frequency difference directly pump the photomixer, a single-tone terahertz signal is generated. This configuration is ideal for producing high-power terahertz signals at a given frequency. For hyperspectral terahertz imaging or spectroscopy, the frequency difference between the optical tones must be varied, and the sample's response measured at each frequency. While this approach achieves high SNR by concentrating all optical pump power on a single frequency, it requires longer measurement times to scan across all frequencies sequentially. (2) Filtered Multi-Tone Optical Pumping: A multi-tone optical beam is used, with adjacent tones having terahertz-range frequency differences. An optical filter isolates and amplifies two tones to create a high-power two-tone beam that pumps the photomixer. This scenario functions similarly to the first but allows unused optical tones to serve additional purposes, such as optical frequency stabilization and frequency/phase locking. (3) Unfiltered Multi-Tone Optical Pumping: In this case, all optical tones are used to generate a multi-tone terahertz signal without filtering. Since the optical pump power is distributed across multiple tones, individual terahertz tones have lower power compared to scenarios (1) and (2). However, all terahertz frequencies are available simultaneously, eliminating the need for frequency tuning. This enables much faster hyperspectral imaging and spectroscopy at the cost of reduced SNR per frequency. Each approach has advantages and trade-offs, making the choice application-dependent based on factors such as SNR, scanning speed, and spectral resolution.

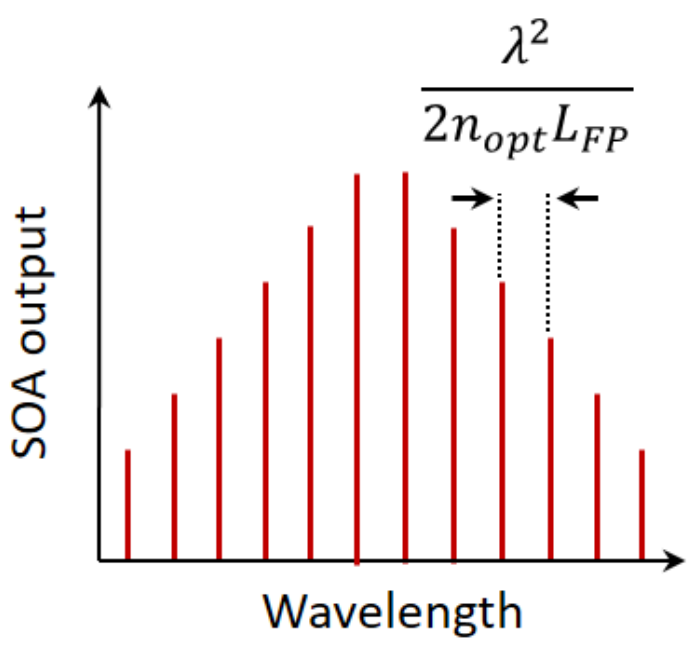

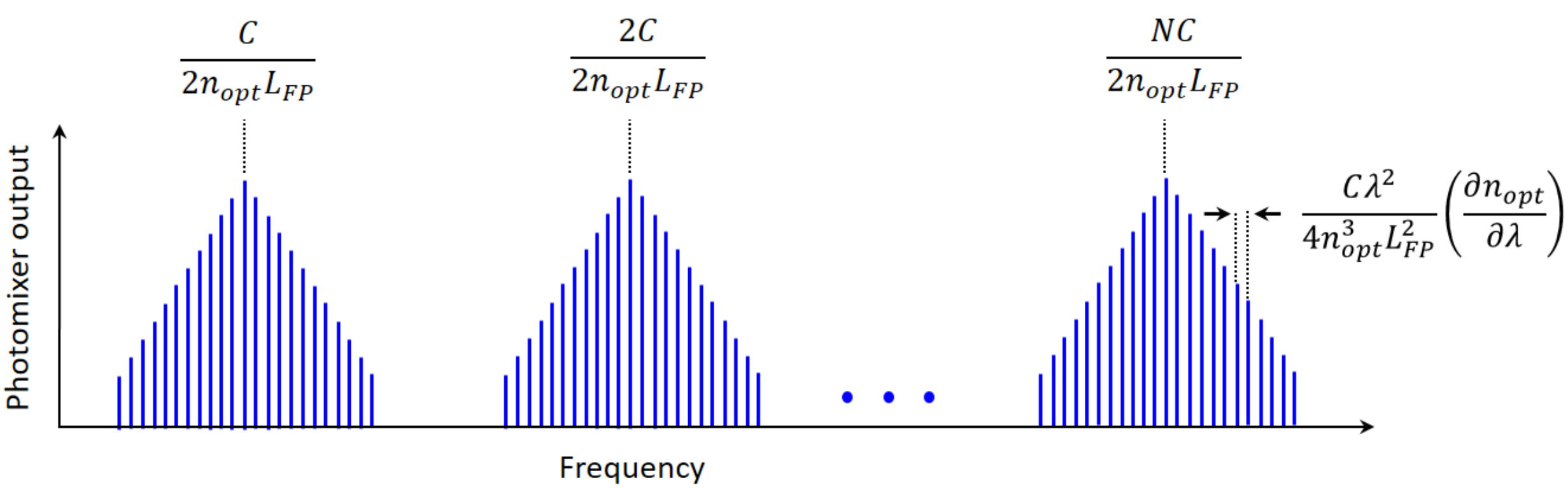

**Fig. S13.** The multi-mode signal generated by the photomixer (blue graph) in response to the multi-mode optical emission (red graph) from the Fabry-Perot laser. The optical tones have a wavelength spacing of $\frac{\lambda^2}{2n_{opt}L_{FP}}$, where $\lambda$ is the optical wavelength, $n_{opt}$ is the effective optical index in the laser cavity, and $L_{FP}$ is the length of the Fabry-Perot cavity. The signal generated by the photomixing process consists of multi-mode tones centered at frequencies of $i\frac{C}{2n_{opt}L_{FP}}$, with a frequency spacing of $\frac{C\lambda^2}{4n_{opt}^3L_{FP}^2}\left(\frac{\partial n_{opt}}{\partial\lambda}\right)$, where $C$ is the speed of light, $i = 1,2,3, \ldots$, and $\frac{\partial n_{opt}}{\partial\lambda}$ represents the dispersion of the optical index inside the laser cavity.

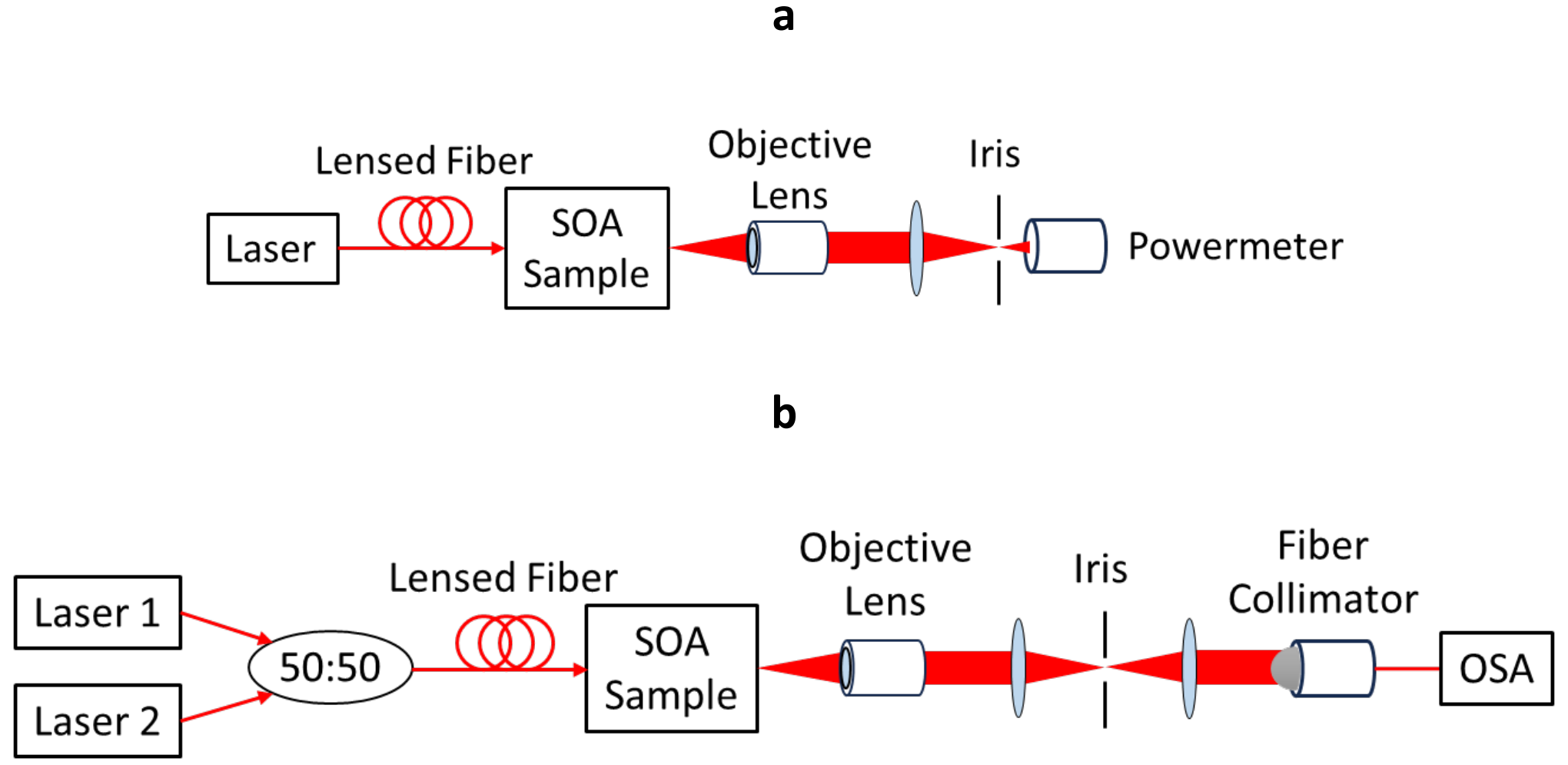

**Fig. S14.** The experimental setups used to characterize the output power and spectral properties of the SOA are shown in **a** and **b**, respectively.

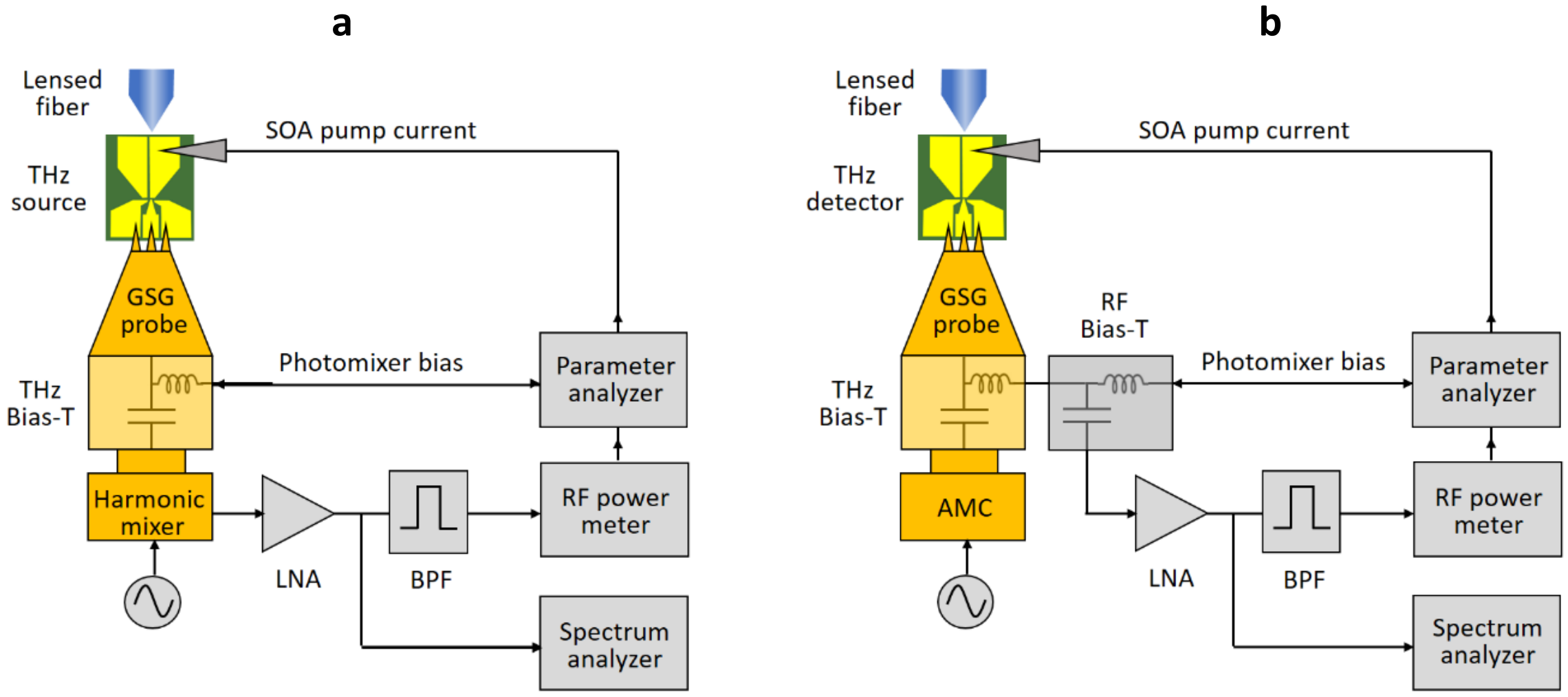

**Fig. S15.** The experimental setups used to characterize the terahertz sources and detectors are shown in **a** and **b**, respectively.

| Reference | Type | Frequency (GHz) | $P_{opt}$ (mW) | $P_{THz}$ (mW) |
|---|---|---|---|---|
| D. Stanze, A. Deninger, A. Roggenbuck, S. Schindler, M. Schlak, B. Sartorius, Compact cw terahertz spectrometer pumped at 1.5 µm wavelength. J Infrared Milli Terahz Waves 32, 225–232 (2011). | Antenna-coupled | 500 | 25 | 0.005 |
| E. Rouvalis, C. C. Renaud, D. G. Moodie, M. J. Robertson, A. J. Seeds, Continuous wave terahertz generation from ultra-fast InP-based photodiodes. IEEE Transactions on Microwave Theory and Techniques 60, 509–517 (2012). | Antenna-coupled | 306 | 48.1 / 38.7 | 0.11 |
| E. Rouvalis, C. C. Renaud, D. G. Moodie, M. J. Robertson, A. J. Seeds, Traveling-wave Uni-Traveling Carrier Photodiodes for continuous wave THz generation. Opt. Express 18, 11105–11110 (2010). | Antenna-coupled | 457 / 255 | 100 /40 | 0.148 / 0.105 |
| J.-M. Wun, H.-Y. Liu, Y.-L. Zeng, S.-D. Yang, C.-L. Pan, C.-B. Huang, J.-W. Shi, Photonic high-power continuous wave THz-wave generation by Using Flip-Chip Packaged Uni-Traveling Carrier Photodiodes and a Femtosecond Optical Pulse Generator. J. Lightwave Technol. 34, 1387–1397 (2016). | Probed | 280 | 130 | 1.04 |
| V. Rymanov, A. Stöhr, S. Dülme, T. Tekin, Triple transit region photodiodes (TTR-PDs) providing high millimeter wave output power. Opt. Express 22, 7550–7558 (2014). | Probed | 110 | 94 | 1.05 |
| I. D. Henning, M. J. Adams, Y. Sun, D. G. Moodie, D. C. Rogers, P. J. Cannard, S. "Jeevan" Dosanjh, M. Skuse, R. J. Firth, Broadband Antenna-Integrated, Edge-Coupled Photomixers for Tuneable Terahertz Sources. IEEE J. Quantum Electron. 46, 1498–1505 (2010). | Antenna-coupled | 410 | 50 | 0.15 |
| A. Wakatsuki, T. Furuta, Y. Muramoto, T. Yoshimatsu, H. Ito, "High-power and broadband sub-terahertz wave generation using a J-band photomixer module with rectangular-waveguide output port" in 2008 33rd International Conference on Infrared, Millimeter and Terahertz Waves (IEEE, 2008), pp. 1–2. | Probed | 350 | 91 | 0.54 |
| H. Ito, T. Yoshimatsu, H. Yamamoto, T. Ishibashi, Widely Frequency Tunable Terahertz-Wave Emitter Integrating Uni-Traveling-Carrier Photodiode and Extended Bowtie Antenna. Appl. Phys. Express 6, 064101 (2013). | Antenna-coupled | 200 / 500 | 47.6 / 47.6 | 0.12 / 0.017 |
| P. Latzel, F. Pavanello, M. Billet, S. Bretin, A. Beck, M. Vanwolleghem, C. Coinon, X. Wallart, E. Peytavit, G. Ducournau, M. Zaknoune, J.-F. Lampin, Generation of mW Level in the 300-GHz Band Using Resonant-Cavity-Enhanced Unitraveling Carrier Photodiodes. IEEE Trans. THz Sci. Technol. 7, 800–807 (2017). | Probed | 300 | 83 | 0.75 |
| J.-M. Wun, C.-H. Lai, N.-W. Chen, J. E. Bowers, J.-W. Shi, Flip-Chip Bonding Packaged THz Photodiode With Broadband High-Power Performance. IEEE Photon. Technol. Lett. 26, 2462–2464 (2014). | Probed | 260 | 162.5 | 0.67 |

| J. Mangeney, A. Merigault, N. Zerounian, P. Crozat, K. Blary, J. F. Lampin, Continuous wave terahertz generation up to 2THz by photomixing on ion-irradiated In0.53Ga0.47As at 1.55μm wavelengths. Applied Physics Letters 91, 241102 (2007). | Antenna-coupled | 500 | 40 | 0.00004 |
|---|---|---|---|---|
| S.-H. Yang, M. Jarrahi, "High-power continuous-wave terahertz generation through plasmonic photomixers" in 2016 IEEE MTT-S International Microwave Symposium (IMS) (IEEE, 2016), pp. 1–4. | Antenna-coupled | 150 | 350 | 0.1 |
| H. Tanoto, J. H. Teng, Q. Y. Wu, M. Sun, Z. N. Chen, S. A. Maier, B. Wang, C. C. Chum, G. Y. Si, A. J. Danner, S. J. Chua, Greatly enhanced continuous-wave terahertz emission by nano-electrodes in a photoconductive photomixer. Nature Photon 6, 121–126 (2012). | Antenna-coupled | 300 / 500 | 90 /90 | 0.09 / 0.011 |
| N. Khiabani, Y. Huang, L. E. Garcia-Munoz, Y.-C. Shen, A. Rivera-Lavado, A Novel Sub-THz Photomixer With Nano-Trapezoidal Electrodes. IEEE Trans. THz Sci. Technol. 4, 501–508 (2014). | Antenna-coupled | 200 / 400 | 30 /30 | 0.0008 / 0.0006 |
| E. Peytavit, S. Lepilliet, F. Hindle, C. Coinon, T. Akalin, G. Ducournau, G. Mouret, J.-F. Lampin, Milliwatt-level output power in the sub-terahertz range generated by photomixing in a GaAs photoconductor. Applied Physics Letters 99, 223508 (2011). | Probed | 305 | 162 | 0.35 |
| E. Peytavit, P. Latzel, F. Pavanello, G. Ducournau, J.-F. Lampin, CW Source Based on Photomixing With Output Power Reaching 1.8 mW at 250 GHz. IEEE Electron Device Lett. 34, 1277–1279 (2013). | Probed | 250 | 270 | 1.8 |
| M. Deumer, S. Nellen, S. Lauck, S. Keyvaninia, S. Berrios, M. Kieper, M. Schell, R. B. Kohlhaas, Ultra-wideband PIN-pd THz emitter with> 5.5 THz bandwidth. Journal of Infrared, Millimeter, and Terahertz Waves 45, 831-840 (2024). | Antenna-coupled | 115 / 300 / 500 | 39.8 mW | 0.88 / 0.062 / 0.012 |
| M. Grzeslo, S. Dülme, S. Clochiatti, T. Neerfeld, T. Haddad, P. Lu, J. Tebart, S. Makhlouf, C. Biurrun-Quel, J. L. Fernández Estévez, J. Lackmann, High saturation photocurrent THz waveguide-type MUTC-photodiodes reaching mW output power within the WR3. 4 band. Optics Express 31, 6484-6498 (2023). | Probed | 240 / 280 | 80 mW | 0.87 / 0.54 |

**Table S1.** The generated power as a function of the optical pump power for photomixer sources demonstrated in the literature over the 100-500 GHz frequency range, including those with free-space optical coupling from an external laser.